\documentclass[sort,compress,3p]{elsarticle}
\usepackage[utf8]{inputenc}
\usepackage{amsmath, amssymb}
\usepackage{mathtools}
\usepackage{color}
\usepackage{array}
\usepackage{pstricks}

\renewcommand{\vec}[1]{\mathbf{#1}}
\newcommand{\vectornorm}[1]{\left|#1\right|}
\newcommand{\compresslist}{%
  \setlength{\itemsep}{1pt}%
  \setlength{\parskip}{0pt}%
  \setlength{\parsep}{0pt}%
}

\begin{document}
\title{Controlling the weights of simulation particles: adaptive particle management using $k$-d trees}
\author[cwi]{Jannis Teunissen\corref{cor1}}
\ead{Jannis.Teunissen@cwi.nl}
\author[cwi,ein]{Ute Ebert}
\address[cwi]{Centrum Wiskunde \& Informatica (CWI), P.O. Box 94079, 1090 GB Amsterdam, The Netherlands}
\address[ein]{Department of Physics, Eindhoven University of Technology, P.O. Box 513, 5600 MB, Eindhoven, The Netherlands}
\cortext[cor1]{Corresponding author}


\begin{abstract}
  In particle simulations, the weights of particles determine how many physical particles they represent.
  Adaptively adjusting these weights can greatly improve the efficiency of the simulation, without creating severe nonphysical artifacts.
  We present a new method for the pairwise merging of particles, in which two particles are combined into one.
  To find particles that are `close' to each other, we use a $k$-d tree data structure.
  With a $k$-d tree, close neighbors can be searched for efficiently, and independently of the mesh used in the simulation.
  The merging can be done in different ways, conserving for example momentum or energy.
  We introduce probabilistic schemes, which set properties for the merged particle using random numbers.
  The effect of various merge schemes on the energy distribution, the momentum distribution and the grid moments is compared.
  We also compare their performance in the simulation of the two-stream instability.
\end{abstract}

\maketitle

\section{Introduction}
Particle-based simulations are widely used, for example to study fluid flows or plasmas.
The physical particles of interest are often not simulated individually, but as groups of particles, called \emph{super-particles} or \emph{macro-particles}.
Most systems contain so many particles that simulating them individually would be very slow or impossible.
And for many macroscopic properties of a system, individual particle behavior is not important.
On the other hand, a sufficient number of particles is required to limit stochastic fluctuations.

The weight of a simulation particle indicates how many physical particles it represents.
Traditionally, particles had a fixed weight~\cite{Birdsall71926, Hockney88}.
More recently, Lapenta and Brackbill~\cite{Lapenta2002317, Lapenta1994213, Lapenta1995139}, Assous et al.~\cite{Assous2003550}, Welch et al.~\cite{Welch2007143}
and others have introduced methods that adapt the weight of particles during a simulation.
As discussed in~\cite{Assous2003550}, adaptive methods have significant advantages if:
\begin{enumerate}
\item Many new particles are created in the simulation. Adaptive re-weighting is required to limit the total number of particles.
  Examples can be found in~\cite{Chanrion2008} and~\cite{Li20121020}.
\item The system has a multiscale nature.
  In some regions more macro-particles are required, especially if some type of mesh refinement is employed, see for example~\cite{lapentaArXiv}.
\item Control is needed over the number of particles per cell, for example to limit stochastic noise to a realistic value.
\end{enumerate}
Our motivation for investigating the adaptive creation of super-particles originated from the simulation of streamer discharges,
as these discharges have both a multiscale nature and strong source terms~\cite{Li20121020, sungrl}.

When changing the weights of simulation particles, the goal is to reduce the number of simulation particles while not altering the physical evolution of the simulated system.
Most methods operate on a single grid cell at a time; arguments for this approach are given in~\cite{Lapenta1995139}.
There are different ways to change the number of particles.
One option is to merge two (or sometimes three) particles, to form particles with higher weights.
Reversely, splitting can be performed to reduce weights.
Another option is to replace all the particles in a cell by a new set of particles, with different weights.
We will use the name `adaptive particle management', introduced in~\cite{Welch2007143}, for all such algorithms.

We present a technique for the merging of particles, that extends earlier work of Lapenta~\cite{Lapenta2002317}.
This method can operate independently of the mesh, and in any space dimension.
The main idea is to store the particle coordinates (typically position and velocity) in a $k$-d tree.
A $k$-d tree is a space partitioning data structure that given $N$ points enables searching for neighbors in $O(\log N)$ time~\cite{Bentley1975}.
We can then efficiently locate pairs of particles with similar coordinates, and these pairs can be merged.
Because the merged particles are similar, the total distribution of particles is not significantly altered.

In section \ref{sec:general}, we briefly discuss the general principles of particle management and $k$-d trees.
The implementation of the new particle management algorithm is discussed in section \ref{sec:method},
where we also introduce different ways to merge particles, which we call `merge schemes'.
In section \ref{sec:numtests}, we compare how the merging of particles affects the particle distribution function for two test distributions.
We also study the effect on the grid moments, such as the particle density, and compare different ways of constructing a $k$-d tree.
As a more practical example, we show how the different merge schemes affect the evolution of the two-stream instability.

\section{Adaptive particle management and $k$-d trees}
\label{sec:general}
As stated in the introduction, it is typically impossible to simulate all the physical particles in a system individually.
Therefore super-particles are used, representing multiple physical particles.
Often, the simulation can run faster or give more accurate results if the weight of these super-particles is controlled adaptively.
Different names have been introduced for these algorithms:
`adaptive particle management'~\cite{Welch2007143},
`control of the number of particles'~\cite{Lapenta1995139},
`particle coalescence'~\cite{Assous2003550},
`particle resampling'~\cite{Chanrion2008},
`particle remapping'~\cite{moss2006},
`particle rezoning'~\cite{Lapenta2002317},
`(particle) number reduction method'~\cite{Shon2001322}
and probably others.
There seem to be many independent findings, with independent names.
We will use the name `adaptive particle management' (APM), introduced in~\cite{Welch2007143}, to describe this class of algorithms.

\subsection{Conservation properties}
If weights of particles are adjusted, then the `microscopic details' of a simulation are changed.
But the relevant macroscopic quantities should be conserved as much as possible.
To specify these macroscopic quantities, we consider a very common type of particle simulation:
the particle in cell (PIC) method, also known as the particle mesh (PM) method~\cite{Hockney88, Birdsall71926}.
In PIC simulations, particles are mapped to moments on a grid.
From the grid moments the fields acting on the particles are computed, and the particles move accordingly.
For example, in an electrostatic code, the charge density is used to compute the electric field.

An APM algorithm typically changes a set of $N_\mathrm{in}$ particles to a new set of $N_\mathrm{out}$ particles.
If the two sets give rise to the same grid moments, they give rise to the same fields.
Therefore, most algorithms are designed to (approximately) conserve the relevant grid moments.

Only conserving the grid moments is not enough, because the dynamics of a system are not fully determined by the fields.
For example, the results of a simulation can be very sensitive to changes in the momentum or energy distribution.
Therefore, some methods try to preserve the shape of these distributions.
More generally, we would like to keep the important aspects of the particle distribution function $f(\vec{x}, \vec{v}, t)$ the same.
The changes to $f(\vec{x}, \vec{v}, t)$ should not be significantly larger than the fluctuations that naturally occur.
For example, in a collision dominated plasma, particles frequently change direction.
Not conserving the momentum distribution in each direction might have little effect on the overall results.
But for a collisionless plasma, a change in the momentum distribution might lead to significant differences.
Similarly, due to the finite number of particles, fluctuations in the local particle density occur naturally.
Therefore, keeping the particle density exactly the same on each grid point might not be necessary, as long as the total number of particles is conserved.

\subsection{Merging and splitting particles}
\label{sec:mergesplit}
\begin{figure}
  \centering
  \begin{minipage}{0.45\textwidth}
    \centering
    \footnotesize
    \input{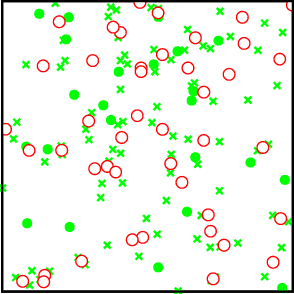}
    \caption{Example showing the merging of particles close in space and velocity (velocity is not shown).
      The particles that were removed after merging are shown as green crosses,
      particles that were not merged as green filled circles,
      and the newly formed merged particles as red empty circles.
      The latter have weight 2, the rest weight 1.
    }
    \label{fig:part_bef_after}
  \end{minipage}
  \hspace{0.05\textwidth}
  \begin{minipage}{0.45\textwidth}
    \centering
    \footnotesize
    \input{kdtree.tex}
    \caption{Schematic example of how a $k$-d tree is generated for points in the plane (indicated as black dots).
      At every step (indicated by the numbers), boxes are split in two parts.
      The split is located on a point, that is added to the tree.
      The direction of splitting alternates between vertical and horizontal.
    }
    \label{fig:kdtree_working}
  \end{minipage}
\end{figure}

A set of $N_\mathrm{in}$ particles can be transformed to a new set of $N_\mathrm{out}$ particles in many ways.
If $N_\mathrm{in} > N_\mathrm{out}$, we use a pairwise coalescence algorithm, that merges two particles into a single new one.
Compared to algorithms that transform multiple particles at the same time, pairwise coalescence has two advantages.
First, it is a more local operation, because only closest neighbors in phase space are selected.
This ensures that the distribution of particles is not changed very much.
Second, it involves fewer degrees of freedom, which makes it simpler to set the properties for new particles.
The pairwise coalescence of particles is illustrated in figure~\ref{fig:part_bef_after}.

In $D$ dimensions, the momentum $\vec{p}$ of the new particle has $D$ degrees of freedom.
Imposing momentum and energy conservation puts $D+1$ constraints on $\vec{p}$.
Therefore, it is in general not possible to conserve both energy and momentum in pairwise coalescence.
This means that there is no single best way to merge particles, as different applications require the conservation of different properties.
We consider several coalescence schemes, which are discussed in section~\ref{sec:mergeschemes}.

The situation would be very different if $N_\mathrm{in}$ particles are merged at the same time to form multiple new particles.
We still have $D+1$ constraints, but now $D \cdot N_\mathrm{out}$ degrees of freedom in the momenta of the $N_\mathrm{out}$ new particles.
The system is under-determined, and additional information about the particles has to be used.
This leads to more complicated algorithms, see for example~\cite{Assous2003550, Welch2007143}.

If $N_\mathrm{in} < N_\mathrm{out}$, particles have to be split.
Several methods for particle splitting have been compared by Lapenta in~\cite{Lapenta2002317}.
As shown there, choosing the right splitting method can be important, depending on the type of simulation.
Here, we will not consider this problem in detail, as our focus is on the merging of particles.
A simple strategy is to split single particles into two new ones with the same properties, but half the weight.
This can be viable if the simulation includes random collisions, so that the new particles will undergo different collisions and spread out.
If there are no such collisions, the split particles should be separated in position or velocity or both.

\subsection{$k$-d trees}
\label{sec:kdtrees}
To locate particles with similar coordinates we use a $k$-d tree~\cite{Bentley1975}, which is a space partitioning data structure.
A $k$-d tree can be used to organize a set of points in a $k$-dimensional space, for any $k\geq1$.
The tree consists of nodes, that contain data (the coordinates of one of the points) and links to at most two `child'-nodes.
The starting point of the tree is the root node, and it contains as many nodes as there are points.

We will briefly explain how such a $k$-d tree can be generated.
To help with the explanation, we let nodes have a \textbf{todo} list, that contains points that need to be processed.
Suppose we have a collection of points in the $(x,y)$ plane.
Initially all points are in the \textbf{todo} list of the root node.
Then the following algorithm, which is illustrated in figure \ref{fig:kdtree_working}, creates the $k$-d tree:
\begin{enumerate}\compresslist
\item Pick a splitting coordinate, either $x$ or $y$. A simple choice is to alternate between them.
\item For each node with a non-empty \textbf{todo} list:
  \begin{enumerate}\compresslist
  \item \label{kdtree:sorting} Sort the particles in the list along the splitting coordinate.
    The particle in the middle of the list is the median.
    If the list contains an even number of particles, pick one of the two middle particles as the median.
  \item The point corresponding to the median is assigned to the node.
  \item The remaining points are moved to the \textbf{todo} lists of (at most) two new child nodes.
    The first one gets the points below the median, the second one those above the median.
  \end{enumerate}
\item If there are still points in \textbf{todo} lists, go back to step one. Otherwise, the tree is completed.
\end{enumerate}
In $k$ dimensions, the only difference would be that there are now $k$ choices for the splitting coordinate.
The computational complexity of creating a $k$-d tree like this is $O(N \log^2 N)$, with $N$ the number of points in the tree.
This can be reduced to $O(N \log N)$ if a linear-time median finding algorithm is used instead of sorting at step~\ref{kdtree:sorting}.
Searching for the nearest neighbor to a location $\vec{r}$ can be done in $O(\log N)$ time.
The basic idea is to first traverse the tree down from the root node, at each step selecting that side of the tree that $\vec{r}$ lies in.
(If $\vec{r}$ happens to lie exactly on a splitting plane, it is a matter of convention which side to pick.)
During the search, the closest neighbor found so far is stored.
Then going upward in the tree, at every step determine whether a closer neighbor could lie on the other side of the splitting plane.
If so, also traverse that other part of the tree down (but only where it can contain a closer neighbor).
Typically, only a small number of these extra traverses is required.
When the algorithm ends up at the root node again, the overall closest neighbor is found.

For the numerical tests presented in section~\ref{sec:numtests}, we have used the Fortran 90 version of the \texttt{KDTREE2}~\cite{Kennel8067K} library.

\section{Implementation}
\label{sec:method}
We will discuss the implementation of our adaptive particle management algorithm in section~\ref{sec:implementationapm}.
Different schemes that can be used for particle merging are given in section~\ref{sec:mergeschemes}.

\subsection{Adaptive particle management algorithm}
\label{sec:implementationapm}
Suppose that we have particles with coordinates $\vec{x}_i$, $\vec{v}_i$ and weights $w_i$.
Furthermore, assume there is some function $W_\mathrm{opt}(i)$ that gives the user-determined optimal weight for particle $i$.
Then the APM algorithm works as follows:

\begin{enumerate}\compresslist
\item \textit{Create a list \textbf{merge} with all the particles for which $w_i < \tfrac{2}{3}  W_\mathrm{opt}(i)$, sorted by $w_i/W_\mathrm{opt}(i)$ from low to high.
    Create a list \textbf{split} with particles for which $w_i > \tfrac{3}{2}  W_\mathrm{opt}(i)$.}

  The function $W_\mathrm{opt}(i)$ gives the desired weight for particle $i$.
  The factors $\tfrac{2}{3}$ and $\tfrac{3}{2}$ ensure that merged particles are not directly split again, and vice versa.
  A good choice of $W_\mathrm{opt}(i)$ will often depend on the application.
  We typically want to keep the number of particles per cell close to a desired value $N_\mathrm{ppc}$,
  and use $W_\mathrm{opt}(i) = \max\left(1, N_\mathrm{phys}(i)/ N_\mathrm{ppc}\right)$.
  Here $N_\mathrm{phys}(i)$ denotes the number of physical particles in the cell of particle $i$.
\item \textit{For the particles in \textbf{merge}:}
  \begin{enumerate}\compresslist
  \item \textit{Create a $k$-d tree with the (transformed) coordinates of the particles as input.}

    We construct the $k$-d trees in two ways: using the coordinates $(\vec{x}, \lambda_v \vec{v})$
    or using the coordinates $(\vec{x}, \lambda_v \vectornorm{\vec{v}})$, where $\lambda_v$ is a scaling parameter and $\vectornorm{}$ denotes the $L^2$ norm.
    We will refer to them as the `full coordinate $k$-d tree' and the `velocity norm $k$-d tree', and we will denote them
    with a superscript $^{\vec{x},\vec{v}}$ and $^{\vec{x},\vectornorm{\vec{v}}}$, respectively.
    The scaling is necessary because the nearest neighbor search uses the Euclidean distance between points.
    There is some freedom in the choice of $\lambda_v$, which should express the ratio of a typical length divided by a typical velocity.
    With higher values the differences in velocity become more important than the spatial distances.
  \item \textit{Search the nearest neighbor of each particle in the $k$-d tree.
      If the distance between particles $i$ and $j$ is smaller than $d_\mathrm{max}$, merge them.
      Particles should not be merged multiple times during the execution of the algorithm, so mark them inactive.}

    We let $d_\mathrm{max}$ be proportional to the grid spacing $\Delta x$, so particles in finer grids need to be closer to be merged.
    There is no single optimal way to merge two particles.
    Several schemes for merging are discussed below in section \ref{sec:mergeschemes}.
  \end{enumerate}

\item \textit{Split each of the particles in \textbf{split} into two new particles.}

  The new particles have the same position and velocity as the original particle $i$, and weights $w_i/2$ and $(w_i+1)/2$ (both rounded down).
  As was discussed in section \ref{sec:mergesplit}, for some applications a different method should be used.

\end{enumerate}

\subsection{Merge schemes}
\label{sec:mergeschemes}
When two particles are merged, it is generally not possible to conserve both energy and momentum.
Therefore we consider different schemes, that conserve either momentum, energy or other properties.
The performance of these schemes is compared in section \ref{sec:numtests}.
We have not used ternary schemes, that merge three particles into two.
As discussed in~\cite{Lapenta2002317}, such schemes do not necessarily perform better, although they can conserve both momentum and energy.
Furthermore, they are more complicated to construct in 2D or 3D.

When particles $i$ and $j$ are merged, the weight of the new particle is always the sum of the weights $w_\mathrm{new} = w_i + w_j$.
For the new position we consider two choices.
It can be the weighted average $\vec{x}_\mathrm{new} = (w_i  \vec{x}_i + w_j  \vec{x}_j)/(w_i+w_j)$.
It can also be picked randomly as either $\vec{x}_i$ or $\vec{x}_j$, with the probabilities proportional to the weights.
If we take the weighted average, then we introduce a (slight) bias in the spatial distribution.
On the other hand, picking the position randomly increases stochastic fluctuations.
For example, suppose we have a cluster of particles, and particles are being merged until there is only one left.
If we use the weighted average position, then we always end up at the center of mass.
So the spatial distribution of particles has become very different, a single peak at the center.
With the probabilistic method we also end up with a single peak, located at the position of one of the original particles.
But now the probability of ending up at particle $i$ is proportional to $w_i$.
Therefore, the `average' spatial distribution has the same shape as before the merging.

Below we list several schemes for picking a new velocity $\vec{v}_\mathrm{new}$.
For convenience of notation, let
\begin{align*}
  \vec{v}_\mathrm{avg} &= (w_i  \vec{v}_i + w_j  \vec{v}_j) / (w_i + w_j),\\
  v^2_\mathrm{avg} &= (w_i  \vectornorm{\vec{v}_i}^2 + w_j  \vectornorm{\vec{v}_j}^2) / (w_i + w_j),
\end{align*}
so $\vec{v}_\mathrm{avg}$ is the weighted average velocity and $v^2_\mathrm{avg}$ is the weighted square norm of the velocity.
The schemes are indicated by the following symbols:

\begin{itemize}
\item[p:] Conserve momentum strictly by taking $\vec{v}_\mathrm{new} = \vec{v}_\mathrm{avg}$.
  Because $\vectornorm{\vec{v}_\mathrm{avg}}^2 \leq v^2_\mathrm{avg}$, the kinetic energy is reduced by an amount
  $\tfrac{1}{2} m w_\mathrm{new}  \left(v^2_\mathrm{avg} - \vectornorm{\vec{v}_\mathrm{avg}}^2\right)$, where $m$ is the mass of a particle with weight one.

\item[$\varepsilon$:] Conserve energy strictly by taking $\vec{v}_\mathrm{new} = \sqrt{v^2_\mathrm{avg}} \cdot \vec{\hat{v}}_\mathrm{avg}$ (the hat denotes a unit vector).
  Because the energy is kept the same, the momentum increases by
  $m w_\mathrm{new}\left(\sqrt{v^2_\mathrm{avg}} - \vectornorm{\vec{v}_\mathrm{avg}}\right) \cdot \vec{\hat{v}}_\mathrm{avg}$.

\item[$\vec{v}_r$:] Conserve both momentum and energy on average, by randomly taking the velocity of one of the particles.
  The probability of choosing the velocity of particle $i$ is proportional to its weight $w_i$.

\item[$\vec{v}_r\varepsilon$:] Randomly take the velocity of one of the particles, but scale it to strictly conserve energy.
  The expected change in momentum is
  $m w_\mathrm{new} \left(\sqrt{v^2_\mathrm{avg}}(w_i\vec{\hat{v}}_{i}+w_j\vec{\hat{v}}_{j})/w_\mathrm{new} - \vec{v}_\mathrm{avg}\right)$,
  which is small if $\vectornorm{\vec{v}_i} \approx \vectornorm{\vec{v}_j}$.
\end{itemize}

Although they are quite simple, we are not aware of other authors that have used schemes with randomness.
It is possible to use multiple schemes, where the choice of scheme depends on the properties of the particles to be merged.

\section{Numerical tests and results}
\label{sec:numtests}
It is difficult to come up with a general test of the performance of an APM algorithm.
The algorithm should not significantly alter the simulation results, compared to a run without super-particles.
At the same time, it should decrease the computational cost as much as possible.
But whether these criteria are met depends on the particular simulation that is performed.
Therefore we first perform tests on a simplified 2D system, using two Gaussian velocity distributions.
In these tests we do not study the time evolution of the system, but focus on the effects of the coalescence algorithm on the particle distribution and on the grid moments.
After that, we investigate how these changes in the particle distribution affect the evolution of a `real' simulation: the two-stream instability in 1D.

\subsection{Effect of the merge schemes on the energy and momentum distribution}
\label{sec:numtest_1}
As stated before, our method works in 1D, 2D, 3D or any other dimension.
In these tests we use a 2D domain with periodic boundary conditions.
The domain consists of $2 \times 2$ cells, each of size $1\times 1$.
(We let lengths and velocities be of order unity, and give them without a unit.)
Initially, particles with weight $1$ are distributed uniformly over the domain.
Then the coalescence algorithm is performed once, with the desired weight of the particles set to $2$.
We compare how the different merge schemes change the momentum and energy distribution.
We also measure their effect on the density, momentum and energy grid moments.

\subsubsection{First test}
\begin{figure}[!t]
  \centering
  \begin{minipage}{0.49\textwidth}
    \centering
    \footnotesize
    \input{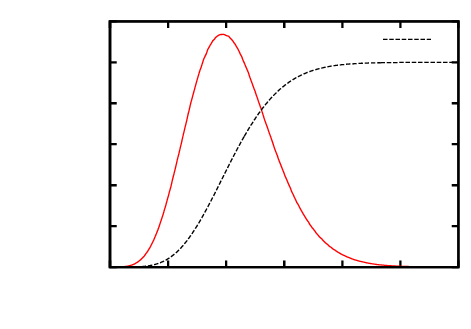}
  \end{minipage}
  \begin{minipage}{0.49\textwidth}
    \centering
    \footnotesize
    \input{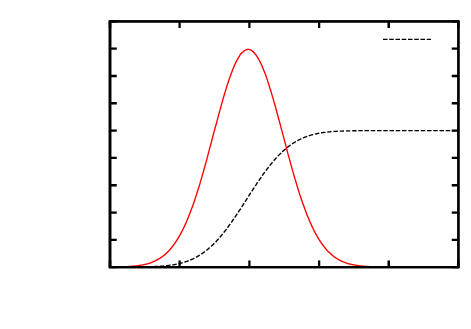}
  \end{minipage}
  \begin{minipage}{0.49\textwidth}
    \centering
    \footnotesize
    \input{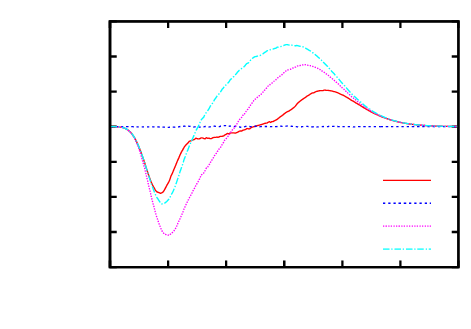}
  \end{minipage}
  \begin{minipage}{0.49\textwidth}
    \centering
    \footnotesize
    \input{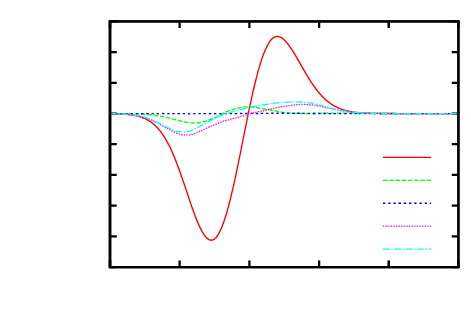}
  \end{minipage}
  \caption{
    Results for the first test.
    Top row: the initial energy (left) and momentum (right) distribution of the particles.
    The integrated or cumulative curves are also shown (dashed).
    Bottom row: the effect of various merge schemes on the cumulative energy (left) and momentum (right) distribution function.
    The schemes are indicated by the following symbols;
    $\varepsilon$: conserve energy,
    p: conserve momentum,
    $\vec{v}_r$: conserve energy and momentum on average,
    $\vec{v}_r\varepsilon$: take velocity from one of the particles at random, scale to conserve energy,
    $^{\vec{x},\vectornorm{\vec{v}}}$: velocity norm $k$-d tree,
    $^{\vec{x},\vec{v}}$: full coordinate $k$-d tree.
  }
  \label{fig:df1_incr_comparison}
\end{figure}

In the first test, there are 400 particles with a Gaussian velocity distribution.
Both components of the velocity have mean $1$ and a standard deviation of $1/4$.
The resulting energy and momentum distribution functions are shown in the top row of figure~\ref{fig:df1_incr_comparison}.
We show the distribution of momentum along the first coordinate, not the total momentum of particles, therefore we label it $x$-momentum.
To convert the velocity of a particle to momentum, we multiply it by the weight of the particle, which represents the mass.

Initially, the particles have weight 1, and a desired weight of 2.
Then the particles are coalesced according to a merge scheme, and the changes in the energy, momentum and density distribution are recorded.
The whole procedure is repeated $10^5$ times for each scheme, using different random numbers, to reduce stochastic fluctuations.
We have used both the velocity norm $k$-d tree (containing $\vec{x}, \lambda_v\vectornorm{\vec{v}}$) and the full coordinate $k$-d tree (containing $\vec{x}, \lambda_v\vec{v}$).
Somewhat arbitrarily we took $\lambda_v = 4/5$, as the mean velocity plus the standard deviation in velocity was $5/4$.

The bottom row of figure~\ref{fig:df1_incr_comparison} shows the effects of the merge schemes on the cumulative energy and momentum distribution function.
The schemes are indicated by the same symbols as in section~\ref{sec:mergeschemes}:
\begin{itemize}\compresslist
\item[p:] conserve momentum
\item[$\varepsilon$:] conserve energy
\item[$\vec{v}_r$:] take velocity of one of the particles at random
\item[$\vec{v}_r\varepsilon$:] take velocity from one of the particles at random, scale to conserve energy
\item[$^{\vec{x},\vectornorm{\vec{v}}}$:] velocity norm $k$-d tree
\item[$^{\vec{x},\vec{v}}$:] full coordinate $k$-d tree
\end{itemize}
Because they are less noisy and reveal trends more clearly, we present cumulative differences
\begin{equation}
  \Delta F(x) = \int_{x_\mathrm{min}}^{x} f_\mathrm{merged}(x') - f_\mathrm{orig}(x')\;dx',
\end{equation}
where $f_\mathrm{orig}(x)$ is the normalized energy or momentum distribution function before merging and $f_\mathrm{merged}(x)$ is the distribution after merging.

The schemes $\varepsilon^{\vec{x},\vectornorm{\vec{v}}}$ and $\vec{v}_r\varepsilon^{\vec{x},\vectornorm{\vec{v}}}$ have the same effect on the energy distribution,
so they are shown together there as $(\vec{v}_r)\varepsilon^{\vec{x},\vectornorm{\vec{v}}}$.
The schemes $\vec{v}_r^{\vec{x},\vectornorm{\vec{v}}}$ and $\vec{v}_r^{\vec{x},\vec{v}}$ are also shown together, as $\vec{v}_r$.
They take the new velocity randomly from one of the original particles.
Therefore, on average, both do not change the shape of the energy and momentum distribution.
The other schemes move particles from the tails of the distribution towards the center.
To see this in the cumulative distribution functions, note that particles get removed where the slope is negative, and are moved to where the slope is positive.
This happens because these schemes take averages, which are more likely to lie towards the center of the distribution.
Results are not shown for the velocity norm $k$-d tree with the momentum conserving scheme, $\text{p}^{\vec{x},\vectornorm{\vec{v}}}$.
This combination leads to large changes in the energy distribution.

For all the merge schemes, on average about $40\%$ of the particles is merged.
The number is below $50\%$ because the $k$-d tree is created only once, in a static way.
When a particle is merged, it is not removed from the tree, but marked as inactive.
So it might later be the nearest neighbor of another particle, that is still to be merged.
In that case, the second particle is not merged, and the algorithm moves on to the next particle.
Another option would be to search for the second closest neighbor, and so on.
But then merging would happen over greater distances towards the end of the algorithm.

Note that even when merging only with the closest neighbor, the order in which the particles are selected for merging can have an effect on the result.
For example, suppose the particles are selected based on their energy, so that particles at lower energies are merged first.
A particle with high energy now has a lower chance of getting merged, because many of its neighbors of lower energy are already merged.
For the same reason, the particle is more likely to be merged with another particle of higher energy.
In the adaptive particle management algorithm introduced in section~\ref{sec:implementationapm},
we therefore sort the particles to be merged by their `relative weights', from low to high,
where `relative weight' means current weight over desired weight.

\subsubsection{Second test}
\label{sec:numtest_2}
\begin{figure}[!t]
  \centering
  \begin{minipage}{0.49\textwidth}
    \centering
    \footnotesize
    \input{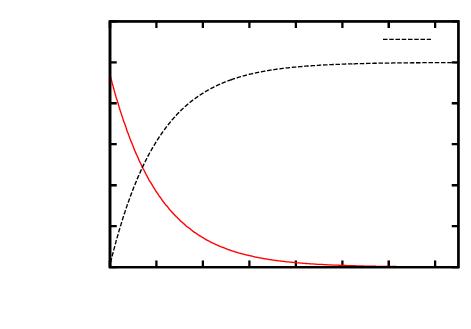}
  \end{minipage}
  \begin{minipage}{0.49\textwidth}
    \centering
    \footnotesize
    \input{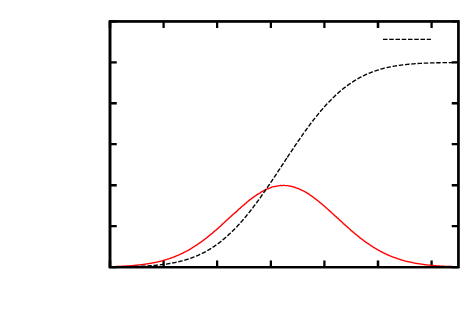}
  \end{minipage}
  \begin{minipage}{0.49\textwidth}
    \centering
    \footnotesize
    \input{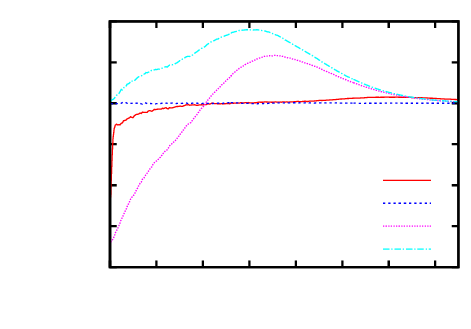}
  \end{minipage}
  \begin{minipage}{0.49\textwidth}
    \centering
    \footnotesize
    \input{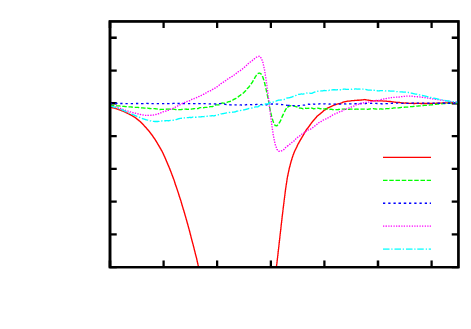}
  \end{minipage}
  \caption{
    Results for the second test.
    Top row: the initial energy (left) and momentum (right) distribution of the particles.
    The integrated or cumulative curves are also shown (dashed).
    Bottom row: the effect of various merge schemes on the cumulative energy (left) and momentum (right) distribution function.
    The legend is the same as for figure~\ref{fig:df1_incr_comparison}.
    In the right figure, the peak for scheme $\varepsilon^{\vec{x},\vectornorm{\vec{v}}}$ is cut off, it extends to $-0.018$.
  }
  \label{fig:df2_incr_comparison}
\end{figure}

The second test is performed in the same way as the first test, but now the particles have a different velocity distribution.
Both components of the velocity have a mean of $1/4$ and a standard deviation of $1$.
The resulting energy and momentum distribution functions are shown in the top row of figure~\ref{fig:df2_incr_comparison}.
Because it is more isotropic, the second velocity distribution poses a bigger challenge for the merge schemes.
The bottom row of figure~\ref{fig:df2_incr_comparison} shows the effects of the merge schemes on the cumulative energy and momentum distribution function.
Again, the schemes $\vec{v}_r$ perform best, as the other schemes move particles from the tail of the distribution towards the center.
Note that the schemes $\vec{v}_r\varepsilon^{\vec{x},\vectornorm{\vec{v}}}$ and $\varepsilon^{\vec{x},\vec{v}}$ also move particles away from zero momentum.
As for the first test, on average about $40\%$ of the particles is merged.

\subsection{Effect on grid moments}
In many particle simulations, a grid (or mesh) is used.
Grid moments are defined at the grid points, and provide local averages from which the fields acting on the particles can be computed.
For example, the first grid moment gives the particle density, the second the current density or momentum density, the third the energy density and so on.
Particles can be mapped to grid moments in different ways, here we use first order interpolation, also know as cloud-in-cell (CIC)~\cite{Birdsall71926, Hockney88}.

Using the data of the second test, we now look at the effect of the merge schemes on the first three grid moments.
An APM algorithm should not induce large differences in these grid moments.
The mean difference is often zero, because the corresponding quantity is conserved.
Therefore, we also look at the relative standard deviation, or $\sigma / \mu$,
where $\sigma$ is the standard deviation of a random variable with mean $\mu$.
This is a measure of the relative size of fluctuations.
We measure these fluctuations at a single grid point, as they would be correlated for multiple grid points.
In table~\ref{tab:schemetable} the changes in the grid moments are given for various schemes.
The schemes are labeled by the same symbols as before.
In addition, $\vec{x}_r$ indicates that the new position is picked randomly from one of the merged particles.
The bottom part of the table is about cell-by-cell merging, which is discussed in section~\ref{sec:cellbycell}.
The APM fluctuations should be compared to those resulting from advancing the particles in time.
Therefore, the table includes entries that list the effect of taking a timestep $\Delta t$ without any merging.
Since we have included no collisions, the particles simply move with a constant velocity during this timestep.

The average deviation in particle density $\rho$ is zero for all the schemes, because they conserve the total weight of the particles.
Therefore this quantity is not included in table~\ref{tab:schemetable}.
The induced fluctuations in the grid moments can differ by almost an order of magnitude between the schemes.
As expected, conserving momentum reduces the mean energy, and conserving energy increases momentum.
This is especially problematic when the velocity norm $k$-d tree is used.
The mean deviations are then larger than $10\%$.
The full coordinate $k$-d tree in combination with the energy-conserving scheme, $\varepsilon^{\vec{x},\vec{v}}$, gives good results regarding energy and momentum conservation.
Schemes that select the new velocity at random do not lead to systematic differences in the energy and momentum grid moments.
With the $\vec{v}_r^{\vec{x},\vectornorm{\vec{v}}}$ scheme, the fluctuations in momentum can be relatively large.
The $\vec{v}_r^{\vec{x},\vec{v}}$ scheme leads to much smaller fluctuations.
This scheme performs well:
on average it conserves the grid moments and also the shapes of the energy/momentum distribution functions, and it does not create big fluctuations.

Taking the new position at random at one of the original particles ($\vec{x}_r$) increases the fluctuations in particle density.
For all the schemes, the fluctuations in density, momentum or energy are smaller than those resulting from a timestep of $\Delta t = 0.4$.

\begin{table}
  \centering
  {\small
    \begin{tabular*}{0.8\textwidth}{@{\extracolsep{\fill}} c | c c r r r r r}
      Method & $N_\mathrm{merge}$ & $d_\mathrm{avg}$ & $\sigma_{\rho}$ & $\Delta p_x$ & $\sigma_{p_x}$ & $\Delta \varepsilon$ & $\sigma_\varepsilon$ \\
      \hline
      $\Delta t = 0.1$ & -   & -    & $1.6\%$ & $0.0\%$ & $9\%$  & $0.0\%$ & $3.8\%$ \\
      $\Delta t = 0.2$ & -   & -    & $2.9\%$ & $0.0\%$ & $16\%$ & $0.0\%$ & $6.7\%$ \\
      $\Delta t = 0.4$ & -   & -    & $4.9\%$ & $0.0\%$ & $25\%$ & $0.0\%$ & $9.4\%$ \\
      \hline
      $\varepsilon^{\vec{x},\vectornorm{\vec{v}}}$          & 39\% & 0.16 & $0.3\%$ & $12\%$ & $16\%$ & $0.0\%$ & $0.8\%$ \\
      p$^{\vec{x},\vectornorm{\vec{v}}}$                    & 39\% & 0.16 & $0.3\%$ & $0.0\%$ & $4\%$ & $-37\%$ & $5.0\%$ \\
      $\vec{v}_r^{\vec{x},\vectornorm{\vec{v}}}$            & 39\% & 0.16 & $0.3\%$ & $0.0\%$ & $24\%$ & $0.0\%$ & $1.2\%$ \\
      $\vec{v}_r\varepsilon^{\vec{x},\vectornorm{\vec{v}}}$ & 39\% & 0.16 & $0.3\%$ & $0.4\%$ & $25\%$ & $0.0\%$ & $0.8\%$ \\
      $\vec{v}_r\vec{x}_r^{\vec{x},\vectornorm{\vec{v}}}$   & 39\% & 0.16 & $1.0\%$ & $0.0\%$ & $24\%$ & $0.0\%$ & $2.2\%$ \\

      $\varepsilon^{\vec{x},\vec{v}}$          & 40\% & 0.38 & $0.7\%$ & $0.1\%$ & $4\%$ & $0.0\%$ & $1.5\%$ \\
      p$^{\vec{x},\vec{v}}$                      & 40\% & 0.38 & $0.7\%$ & $0.0\%$ & $4\%$ & $-1.2\%$& $1.5\%$ \\
      $\vec{v}_r^{\vec{x},\vec{v}}$            & 40\% & 0.38 & $0.7\%$ & $0.0\%$ & $6\%$ & $0.0\%$ & $2.4\%$ \\
      \hline
      p$^\vec{v}$, cell                 & 40\% & 0.19 & $2.8\%$ & $0.0\%$ & $12\%$ & $-0.9\%$ & $3.8\%$ \\
      $\vec{v}_r^{\vec{x},\vectornorm{\vec{v}}}$, cell       & 39\% & 0.17 & $0.3\%$ & $0.0\%$ & $23\%$ & $0.0\%$ & $1.5\%$ \\
      $\vec{v}_r\vec{x}_r^{\vec{x},\vectornorm{\vec{v}}}$, cell & 39\% & 0.17 & $1.0\%$ & $0.0\%$ & $23\%$ & $0.0\%$ & $2.2\%$\\
      $\varepsilon^{\vec{x},\vec{v}}$, cell        & 38\% & 0.40 & $0.8\%$ & $0.5\%$ & $5\%$ & $0.0\%$ & $1.8\%$
    \end{tabular*}
  }
  \caption{The induced differences and fluctuations in the grid moments by the various merge schemes, using the second test distribution.
    Legend: $N_\mathrm{merge}$ is the fraction of merged particles, and $d_\mathrm{avg}$ is the average distance between merged particles.
    The relative differences in grid moments are indicated by $\Delta p_x$ (momentum) and $\Delta \varepsilon$ (energy),
    and relative standard deviations by $\sigma_{\rho}$ (density), $\sigma_{p_x}$ (momentum) and $\sigma_\varepsilon$ (energy).
    Both are given relative to the mean value.
    The rows starting with $\Delta t$ show the fluctuations in the grid moments resulting from a timestep (no merging).
    The merge schemes are indicated by the following symbols;
    $\varepsilon$: conserve energy,
    p: conserve momentum,
    $\vec{v}_r$: random velocity,
    $\vec{v}_r\varepsilon$: random velocity, scale to conserve energy,
    $\vec{x}_r$: random position,
    $^{\vec{v}}$: $k$-d tree contains only the velocity,
    $^{\vec{x},\vectornorm{\vec{v}}}$: velocity norm $k$-d tree,
    $^{\vec{x},\vec{v}}$: full coordinate $k$-d tree,
    cell: perform the merging cell-by-cell.
  }
  \label{tab:schemetable}
\end{table}

\subsection{Cell-by-cell merging}
\label{sec:cellbycell}
Using $k$-d trees, there is no reason to do cell-by-cell merging.
But because this type of merging is commonly used, we briefly evaluate its effects.
The bottom part of table~\ref{tab:schemetable} shows results for cell-by-cell merging, for various schemes, using the second test distribution.
The notation is the same as before, and a superscript $^{\vec{v}}$ indicates that only the velocity was used in the $k$-d tree, not the position.
The fluctuations are mostly similar if the particles are merged locally (cell-by-cell) instead of globally.
With fewer particles per cell, the differences would be larger though, as close neighbors are more likely to lie in other cells.

\begin{figure}
  \centering
  \begin{minipage}{0.072\textwidth}
    \centering
    \includegraphics[width=1.0\textwidth]{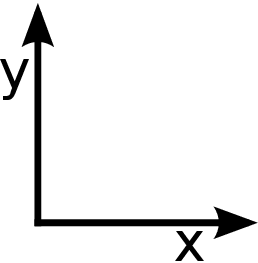}
  \end{minipage}
  \begin{minipage}{0.3\textwidth}
    \centering
    \footnotesize
    \input{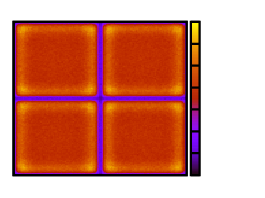}
  \end{minipage}
  \begin{minipage}{0.3\textwidth}
    \centering
    \footnotesize
    \input{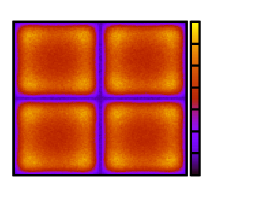}
  \end{minipage}
  \begin{minipage}{0.3\textwidth}
    \centering
    \footnotesize
    \input{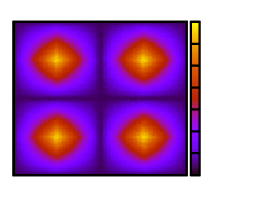}
  \end{minipage}
  \caption{The relative change in particle density as a result of cell-by-cell merging, for the test case described in~\ref{sec:numtest_2}.
    The $2\times 2$ grid structure is clearly visible.
    On the left, close neighbors for merging are searched for using the difference in position and in the norm of the velocity.
    In the middle, all four position and velocity coordinates are used, while on the right only the differences in velocity are considered for merging.
    Especially in the last case, particles are moved to the center of the cell as they are merged, because it is the expectation value of their mean position.
  }
  \label{fig:densdist}
\end{figure}

The average spatial distribution of particles directly after merging is shown in figure~\ref{fig:densdist}.
Only the type of $k$-d tree is important for the effect, because all the shown merge schemes take the average position.
From left to right:
With the velocity norm $k$-d tree the spatial distribution of particles is affected close to the cell boundaries.
With the full coordinate $k$-d tree, the effect is similar as with the velocity norm $k$-d tree.
Using the $k$-d tree that includes only the velocity, particles are moved to the center of the cells.
The spatial distribution is severely affected.
Furthermore, the fluctuations in particle density are higher, as can be seen in table~\ref{tab:schemetable}.
If particles would be merged globally with a $k$-d tree (not cell-by-cell), the spatial distribution would on average be uniform.

Since the density fluctuations due to cell-by-cell merging occur at length scales smaller than the grid resolution,
we do not expect them to have practical consequences for most simulations.
But in some cases, for example if spatial features expand in time, they might become important.

\subsection{Simulation example: the two-stream instability}
Above, we have studied the changes in momentum and energy distribution that arise due to the merging of particles.
But do these changes in the distribution function affect the physical evolution that one wants to study?
The answer will, of course, depend on the type of simulation that is performed.
Here, we consider as an example the simulation of the two-stream instability in one dimension~\cite{bittencourt2004}.
To induce this instability we create two beams of particles, that propagate in opposite directions.
The particles have the same charge, and are neutralized by a background charge density.
A fluctuation in the charge density locally changes the electric field, which affects the beam velocities, which can in turn increase the density fluctuation.

\begin{figure}
  \centering
  \footnotesize
  \begin{tabular}{>{\centering\arraybackslash} m{0.072\textwidth} @{\hspace{0.008\textwidth}}
      >{\centering\arraybackslash} m{0.12\textwidth} @{\hspace{0.008\textwidth}}
      >{\centering\arraybackslash} m{0.12\textwidth} @{\hspace{0.008\textwidth}}
      >{\centering\arraybackslash} m{0.12\textwidth} @{\hspace{0.008\textwidth}}
      >{\centering\arraybackslash} m{0.12\textwidth} @{\hspace{0.008\textwidth}}
      >{\centering\arraybackslash} m{0.12\textwidth} @{\hspace{0.008\textwidth}}
      >{\centering\arraybackslash} m{0.12\textwidth} @{\hspace{0.008\textwidth}}
      >{\centering\arraybackslash} m{0.12\textwidth} @{\hspace{0.008\textwidth}}
    }
    \includegraphics[width=0.072\textwidth]{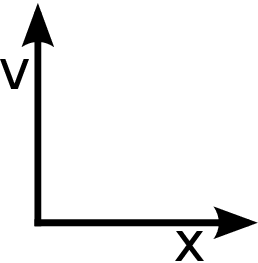} &
    \includegraphics[width=0.12\textwidth]{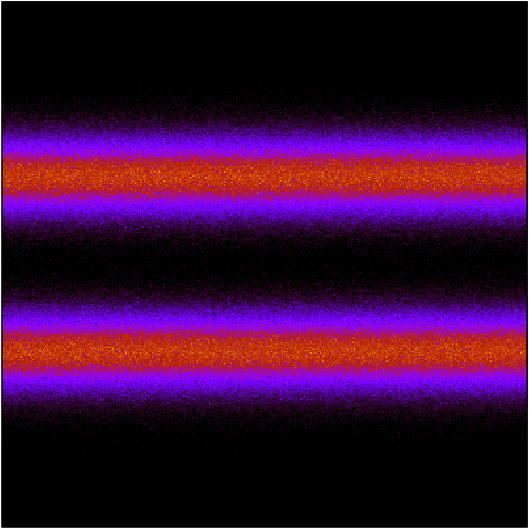} &
    \includegraphics[width=0.12\textwidth]{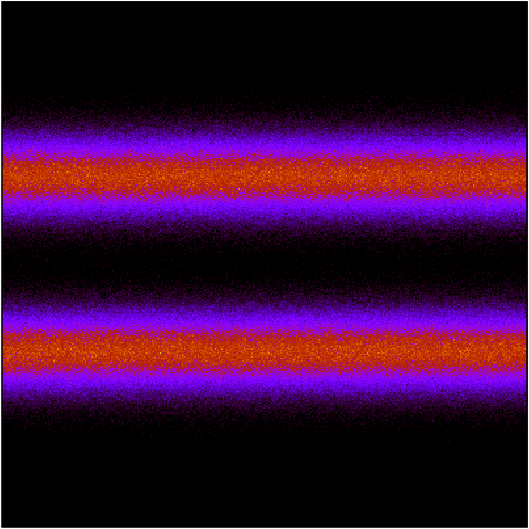} &
    \includegraphics[width=0.12\textwidth]{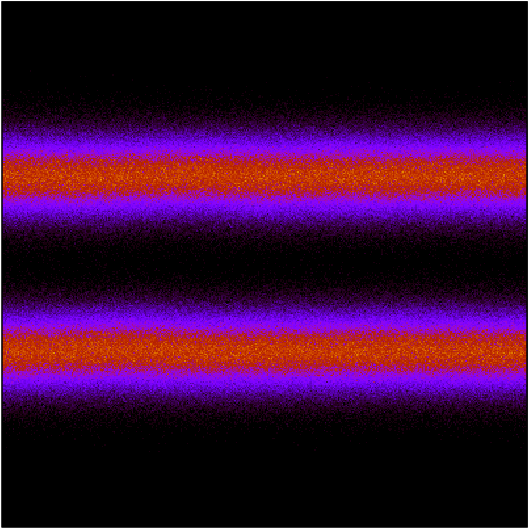} &
    \includegraphics[width=0.12\textwidth]{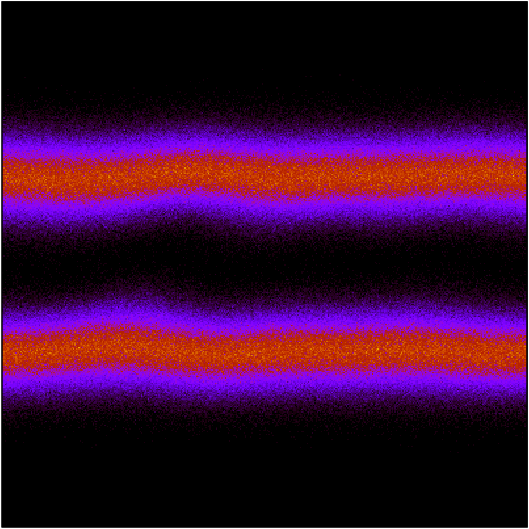} &
    \includegraphics[width=0.12\textwidth]{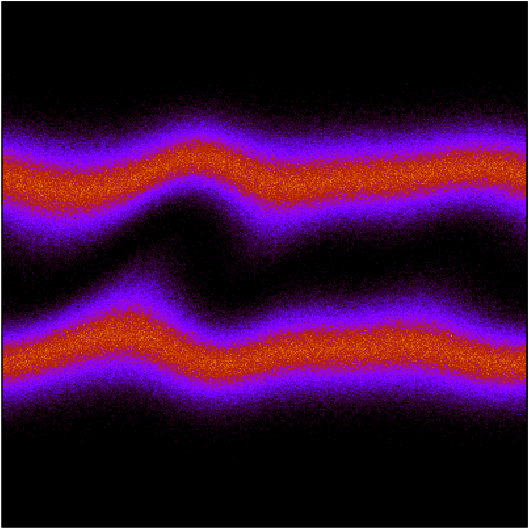} &
    \includegraphics[width=0.12\textwidth]{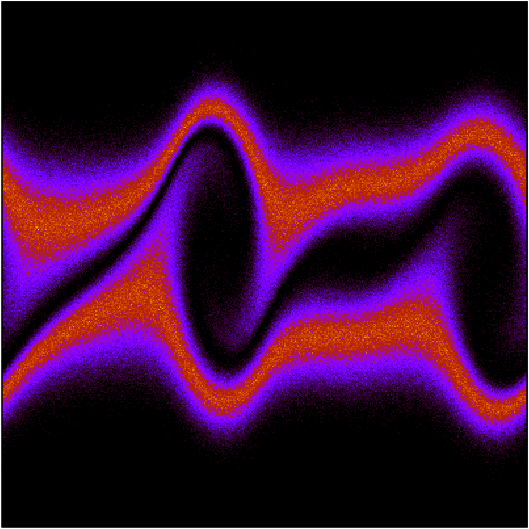} &
    \includegraphics[width=0.12\textwidth]{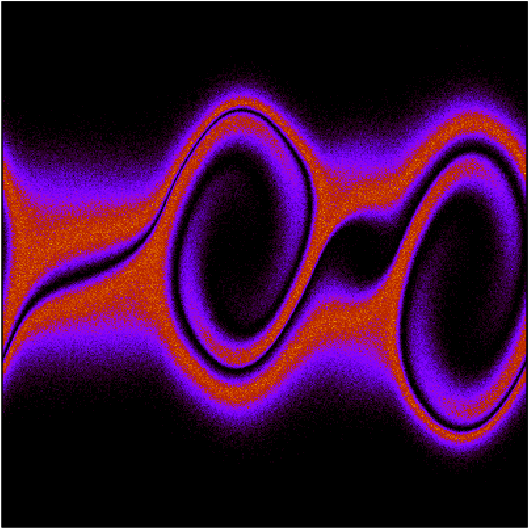} \\
    & $t = 5/\omega_p$ & $10/\omega_p$ & $15/\omega_p$ & $20/\omega_p$ & $25/\omega_p$ & $30/\omega_p$ & $35/\omega_p$
  \end{tabular}
  \caption{The time evolution of the two-stream instability. In each figure, the periodic domain is shown horizontally.
    Vertically, the density in velocity space is shown.
    Initially, the two beams of particles are clearly visible.
    The time $t$ is indicated in inverse plasma frequencies $1/\omega_p$.
    \label{fig:two_stream_example}}
\end{figure}

In figure~\ref{fig:two_stream_example}, an example of the evolution of the two-stream instability is shown.
We use periodic boundary conditions, and there are $N_p = 10^6$ particles per beam.
The particles have a Gaussian velocity distribution, with a standard deviation or thermal velocity $v_\mathrm{th} = 1$ and a drift velocity $v_d = \pm 4$.
The spatial grid consists of $10^3$ points and has length $L = 1$.
To convert the particle line density to a volume density $n$, scaling by some unit of area is required.
We do this in such a way that there are 100 Debye lengths in the domain
\begin{equation}
  \lambda_D = \sqrt{\frac{\epsilon_0 k_B T}{n q^2}} = L/100,
\end{equation}
where $k_B$ is the Boltzmann constant, $\epsilon_0$ the permittivity of vacuum and $q$ the charge of the particles.
Defining the temperature of the particles in a beam as $T = m v_\mathrm{th}^2 / k_B$, with $m$ the mass of the particles,
the plasma frequency is then given by
\begin{equation}
  \omega_p = \sqrt{\frac{n q^2}{m \varepsilon_0}} = 100 L / v_\mathrm{th},
\end{equation}
or simply $\omega_p = 100$ in dimensionless units.
We will give simulation times in inverse plasma frequencies.
Note that when defining the simulation parameters like this, the constants used for the mass and charge of the particles do not matter.

We perform two tests to investigate how the merging of particles affects the physical evolution of the system.
In the first test, merging starts at $t = 5 / \omega_p$, when no instability is yet visible in figure~\ref{fig:two_stream_example}.
In the second test, merging starts later, at $t = 20/\omega_p$, when instabilities have grown to a visible size.
The desired weight of the particles is set to 32, and the merging routine is called five times.
For the merging, we either use the energy conserving scheme ($\varepsilon$), the momentum conserving scheme (p),
the scheme that picks a velocity at random from one of the original particles ($\vec{v}_r$) or the scheme that picks a velocity at random
but conserves energy ($\vec{v}_r\varepsilon$).
The typical distance between particles in space is $\delta x = L/N_p$, while the typical difference in velocity is $\delta v = v_\mathrm{th}$.
We construct the coordinates for the $k$-d tree using $(x, \lambda_v v)$, with $\lambda_v = 10 \delta x / \delta v$.
The value of $\lambda_v$ is adjusted as the number of simulation particles changes.

\begin{figure}
  \centering
  \begin{tabular}{
      >{\centering\arraybackslash} m{0.10\textwidth}
      >{\centering\arraybackslash} m{0.10\textwidth} @{\hspace{0.008\textwidth}}
      >{\centering\arraybackslash} m{0.10\textwidth} @{\hspace{0.008\textwidth}}
      >{\centering\arraybackslash} m{0.10\textwidth} @{\hspace{0.008\textwidth}}
      >{\centering\arraybackslash} m{0.10\textwidth} @{\hspace{0.008\textwidth}}
      >{\centering\arraybackslash} m{0.10\textwidth} @{\hspace{0.008\textwidth}}
      >{\centering\arraybackslash} m{0.10\textwidth} @{\hspace{0.008\textwidth}}
      >{\centering\arraybackslash} m{0.10\textwidth}}
    \footnotesize
    original &
    \includegraphics[width = 0.10\textwidth]{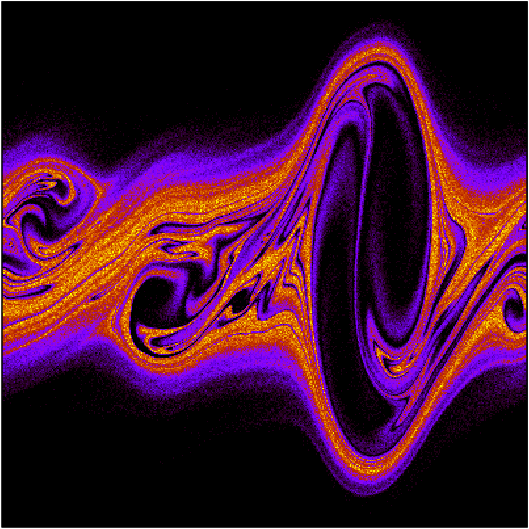} &
    \includegraphics[width = 0.10\textwidth]{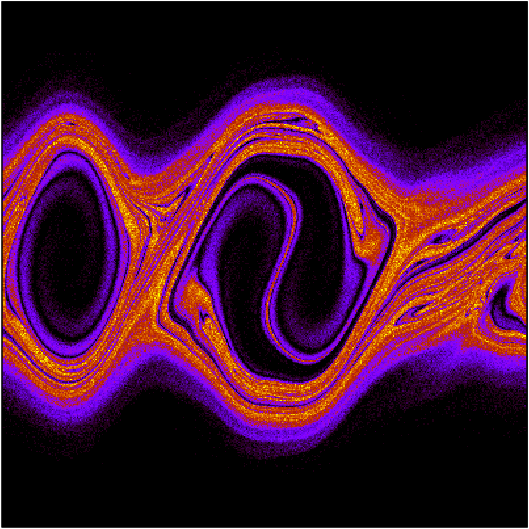} &
    \includegraphics[width = 0.10\textwidth]{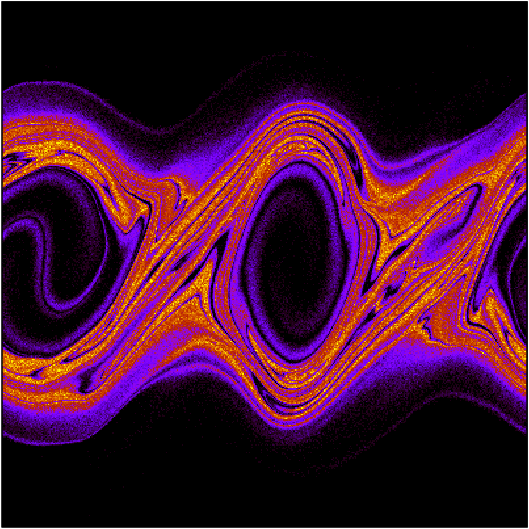} &
    \includegraphics[width = 0.10\textwidth]{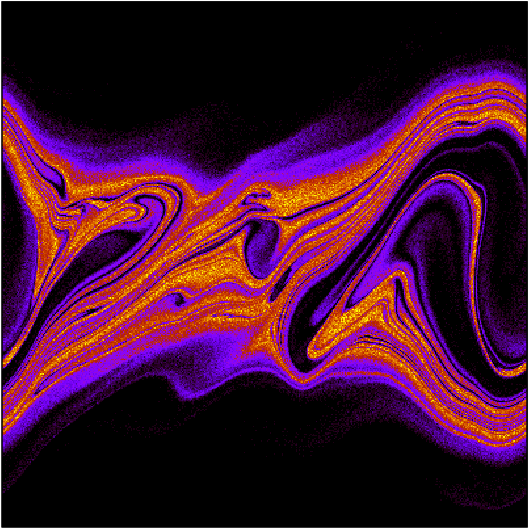} &
    \includegraphics[width = 0.10\textwidth]{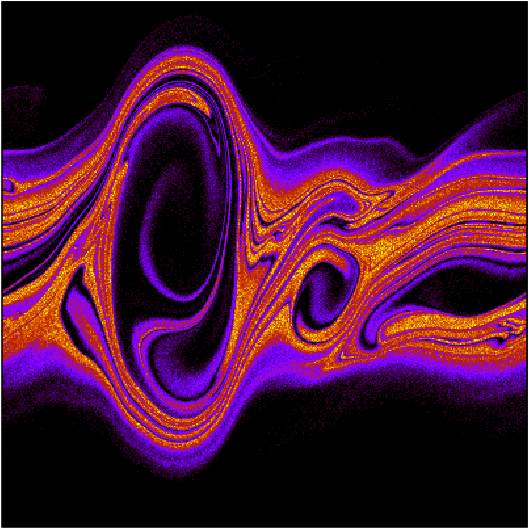} &
    \includegraphics[width = 0.10\textwidth]{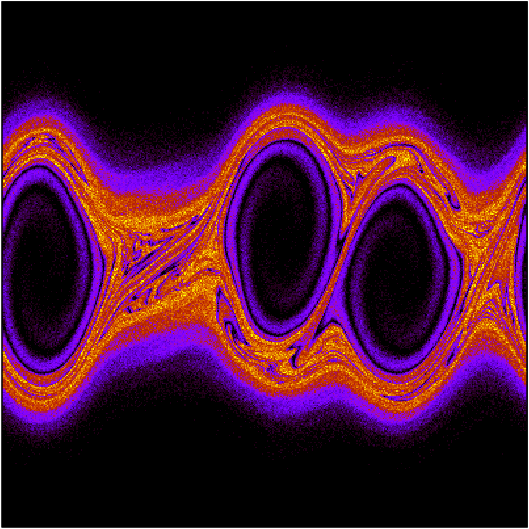} &
    \includegraphics[width = 0.10\textwidth]{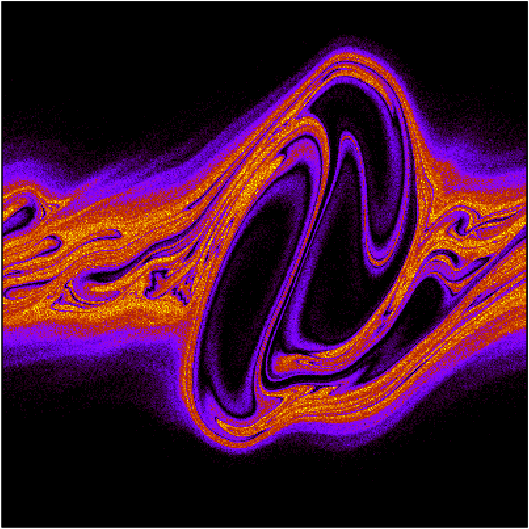} \\
    ($\varepsilon$) &
    \includegraphics[width = 0.10\textwidth]{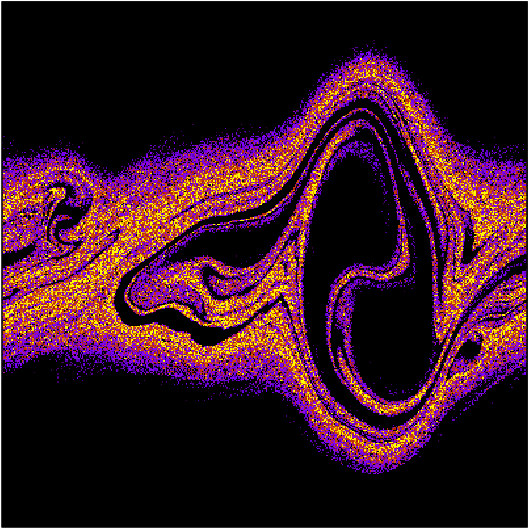} &
    \includegraphics[width = 0.10\textwidth]{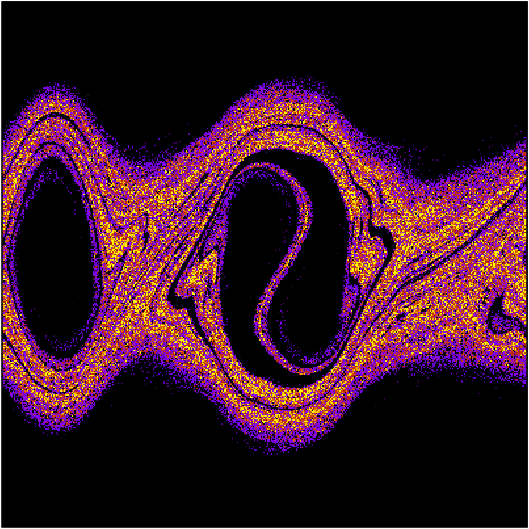} &
    \includegraphics[width = 0.10\textwidth]{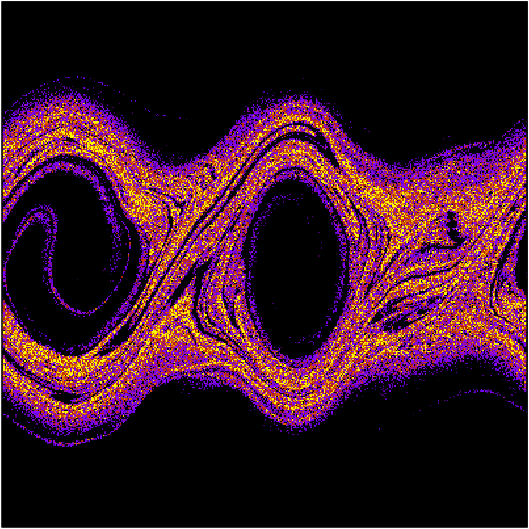} &
    \includegraphics[width = 0.10\textwidth]{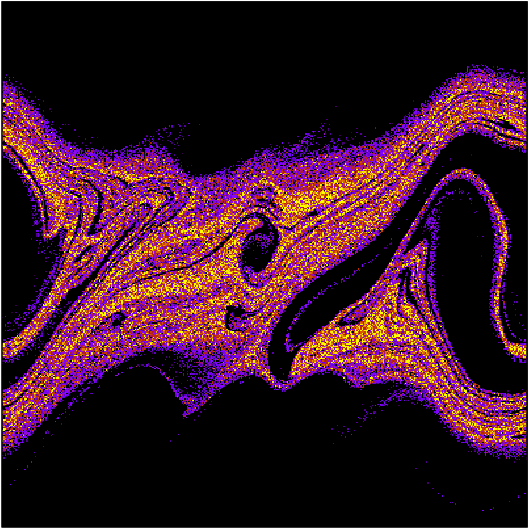} &
    \includegraphics[width = 0.10\textwidth]{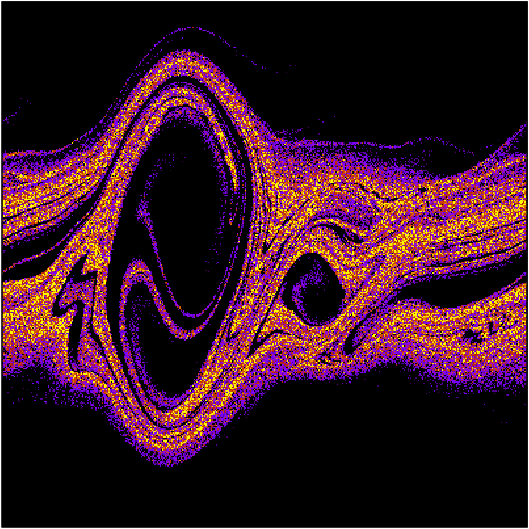} &
    \includegraphics[width = 0.10\textwidth]{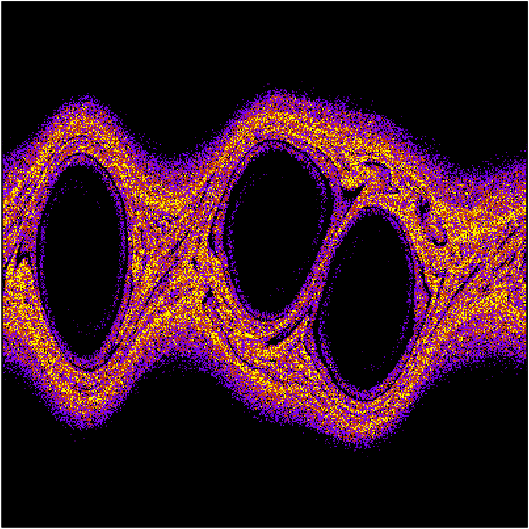} &
    \includegraphics[width = 0.10\textwidth]{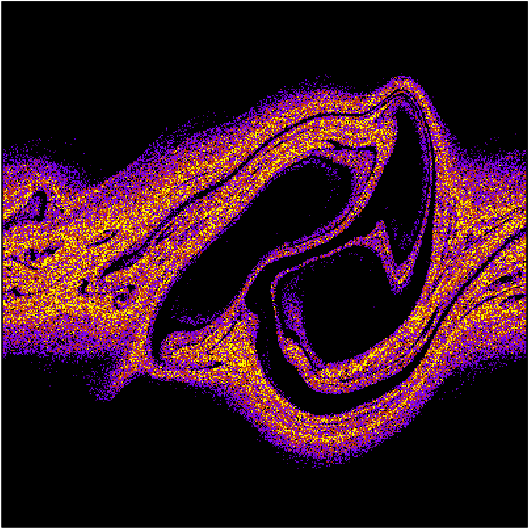} \\
    (p) &
    \includegraphics[width = 0.10\textwidth]{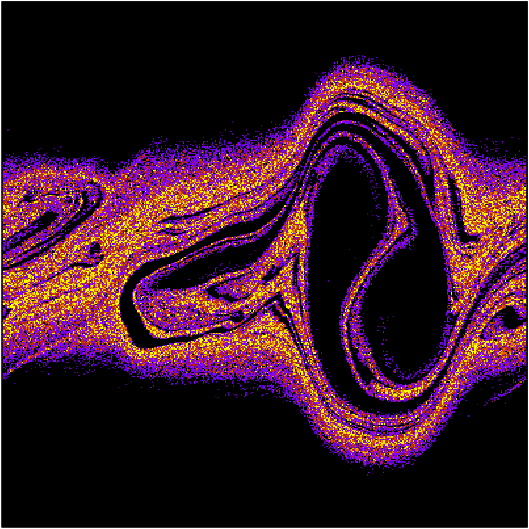} &
    \includegraphics[width = 0.10\textwidth]{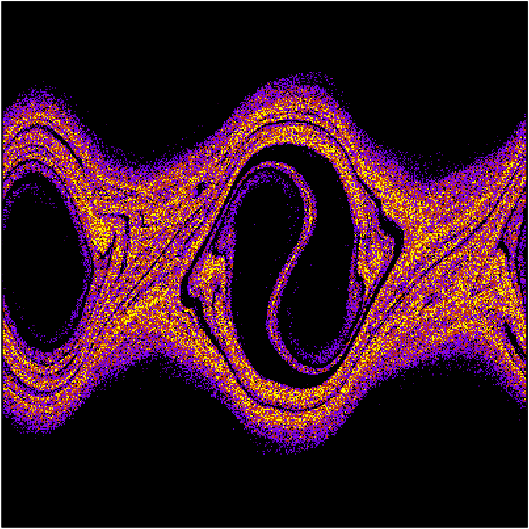} &
    \includegraphics[width = 0.10\textwidth]{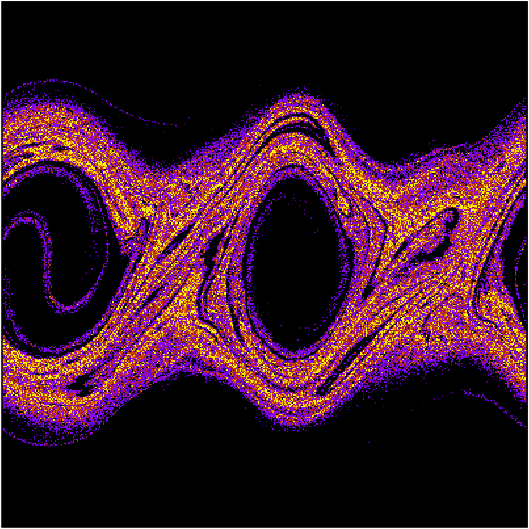} &
    \includegraphics[width = 0.10\textwidth]{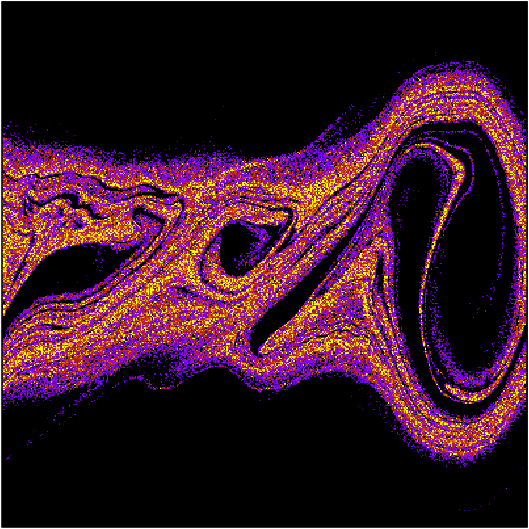} &
    \includegraphics[width = 0.10\textwidth]{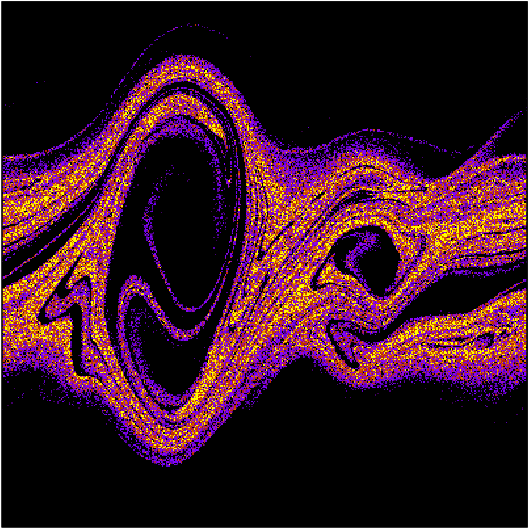} &
    \includegraphics[width = 0.10\textwidth]{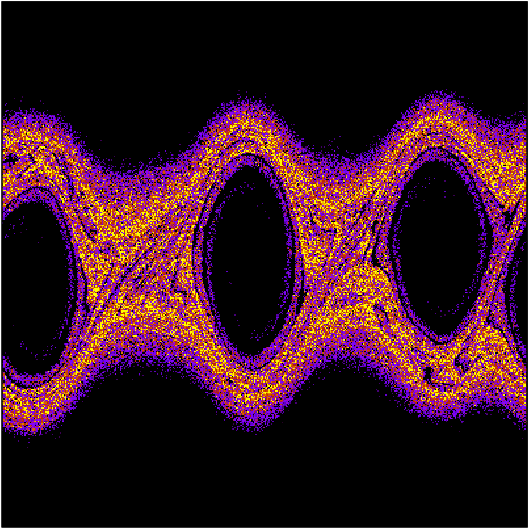} &
    \includegraphics[width = 0.10\textwidth]{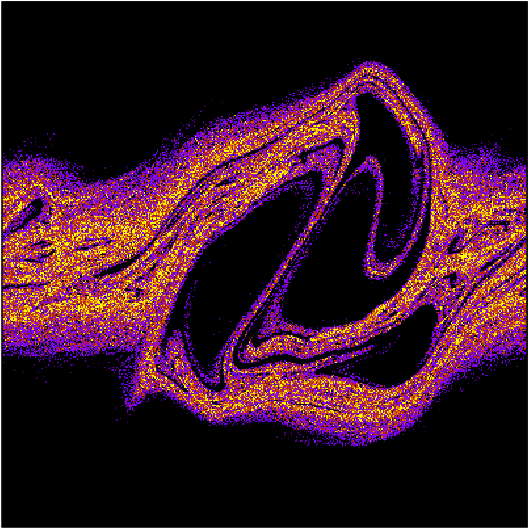} \\
    ($\vec{v}_r$) &
    \includegraphics[width = 0.10\textwidth]{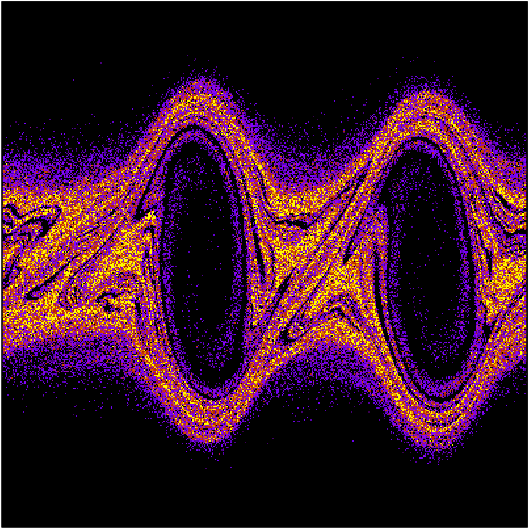} &
    \includegraphics[width = 0.10\textwidth]{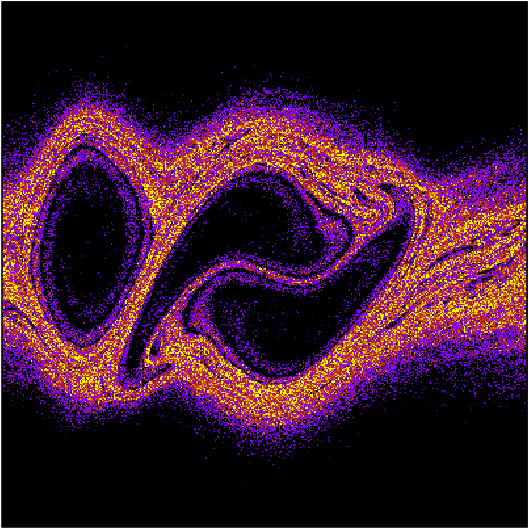} &
    \includegraphics[width = 0.10\textwidth]{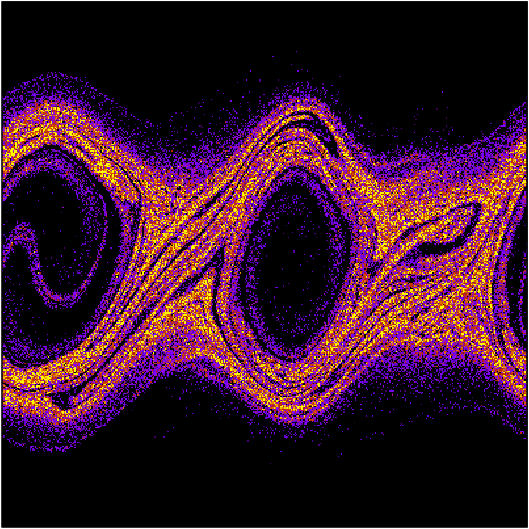} &
    \includegraphics[width = 0.10\textwidth]{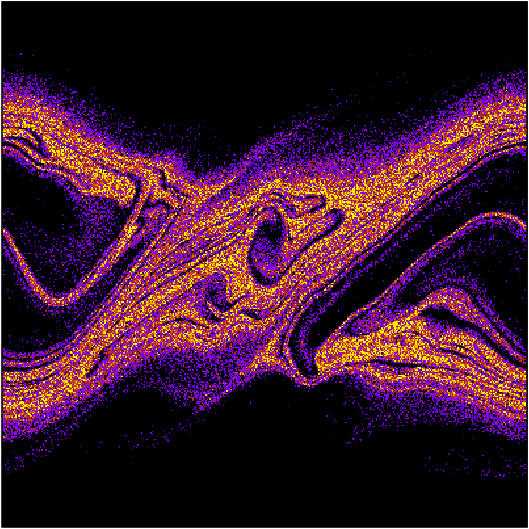} &
    \includegraphics[width = 0.10\textwidth]{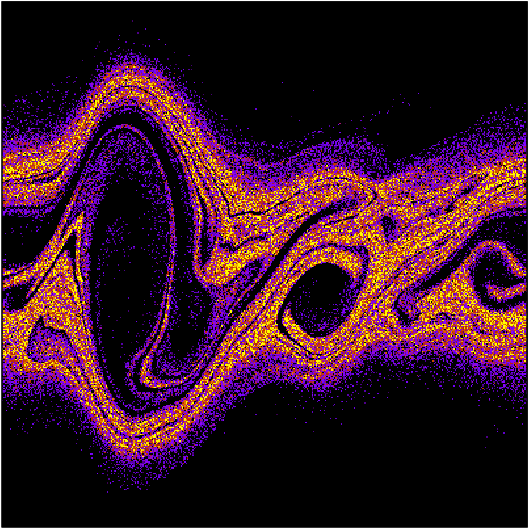} &
    \includegraphics[width = 0.10\textwidth]{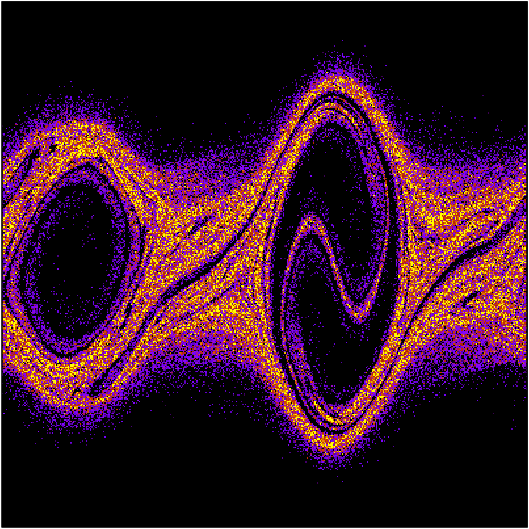} &
    \includegraphics[width = 0.10\textwidth]{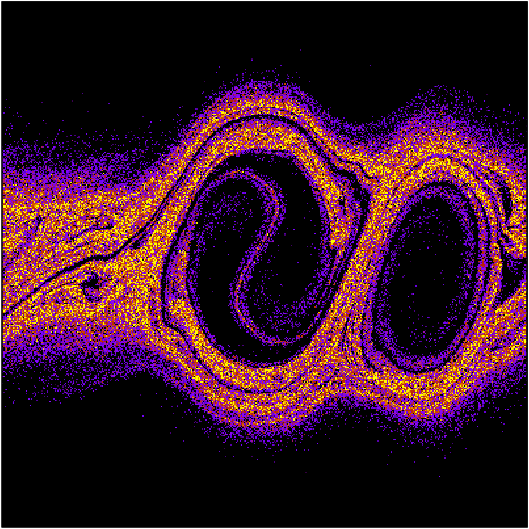} \\
    ($\vec{v}_r\varepsilon$) &
    \includegraphics[width = 0.10\textwidth]{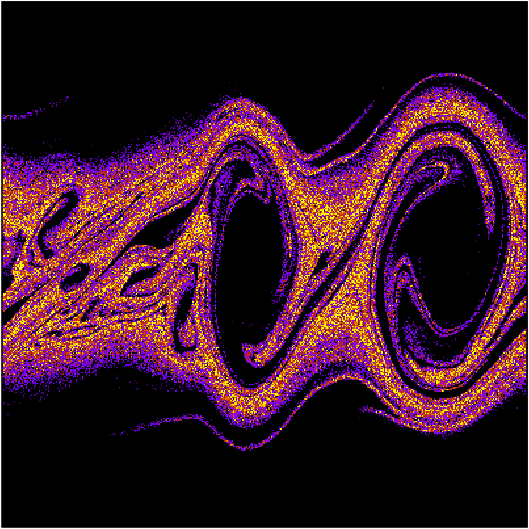} &
    \includegraphics[width = 0.10\textwidth]{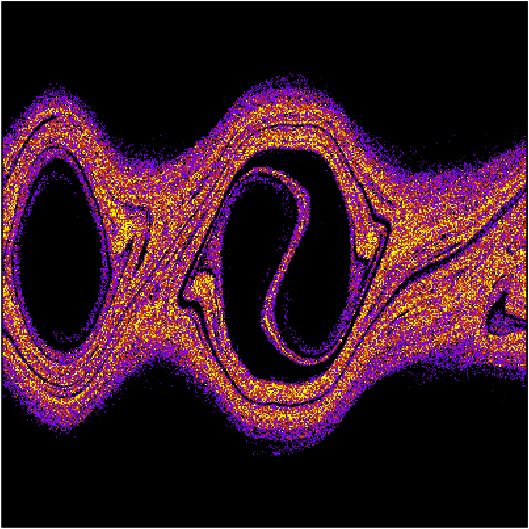} &
    \includegraphics[width = 0.10\textwidth]{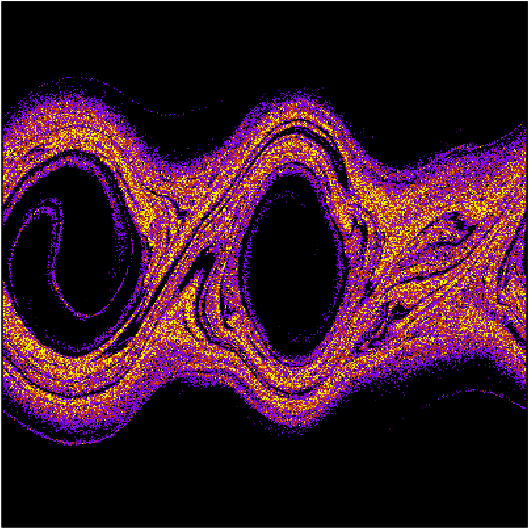} &
    \includegraphics[width = 0.10\textwidth]{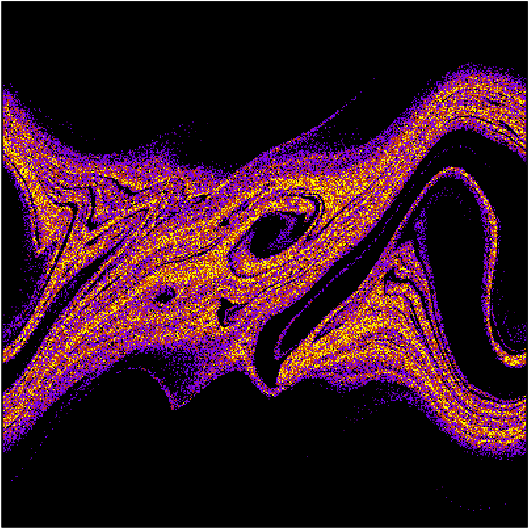} &
    \includegraphics[width = 0.10\textwidth]{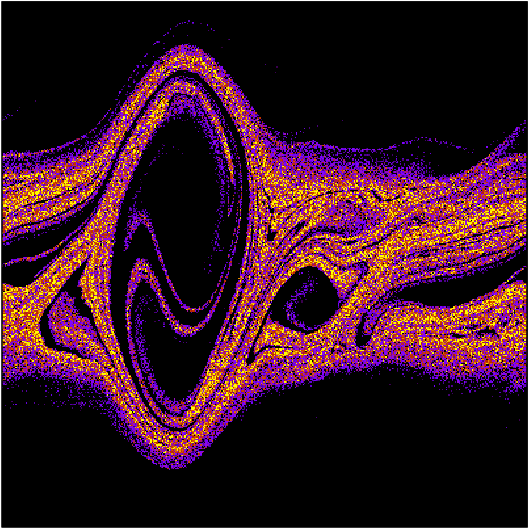} &
    \includegraphics[width = 0.10\textwidth]{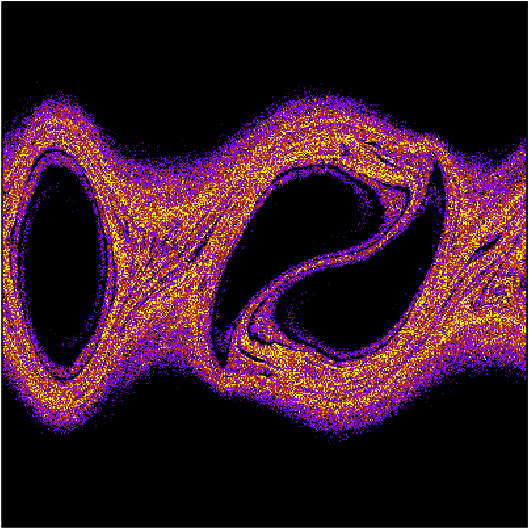} &
    \includegraphics[width = 0.10\textwidth]{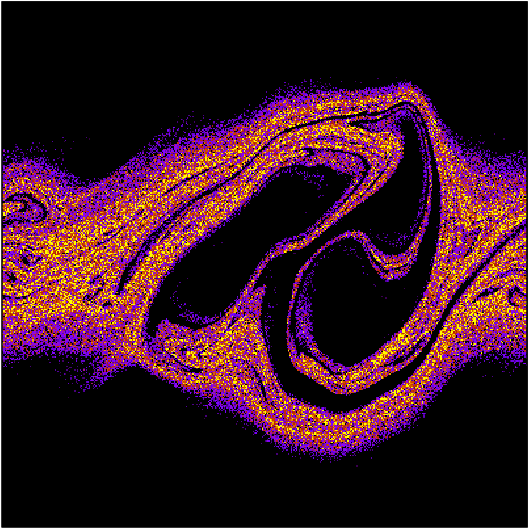} \\
    & run 1 & run 2 & run 3 & run 4 & run 5 & run 6 & run 7
  \end{tabular}
  \caption{Simulation results of the two-stream instability, showing $x,v$ curves at $t = 60/\omega_p$ (see also figure~\ref{fig:two_stream_example}).
    The top row shows the original simulation with $10^6$ particles per beam.
    The other rows show results where repeated merging took place at $t = 5 / \omega_p$, reducing the particle number to about $8\cdot 10^4$ particles per beam.
    The columns show different runs, differing in the initial state of the pseudorandom number generator.
    The $\varepsilon$ scheme seems to give results closest to the original evolution.
    \label{fig:two_str_merging_1}}
\end{figure}

\begin{figure}
  \centering
  \begin{tabular}{
      >{\centering\arraybackslash} m{0.10\textwidth}
      >{\centering\arraybackslash} m{0.10\textwidth} @{\hspace{0.008\textwidth}}
      >{\centering\arraybackslash} m{0.10\textwidth} @{\hspace{0.008\textwidth}}
      >{\centering\arraybackslash} m{0.10\textwidth} @{\hspace{0.008\textwidth}}
      >{\centering\arraybackslash} m{0.10\textwidth} @{\hspace{0.008\textwidth}}
      >{\centering\arraybackslash} m{0.10\textwidth} @{\hspace{0.008\textwidth}}
      >{\centering\arraybackslash} m{0.10\textwidth} @{\hspace{0.008\textwidth}}
      >{\centering\arraybackslash} m{0.10\textwidth}}
    \footnotesize
    original &
    \includegraphics[width = 0.10\textwidth]{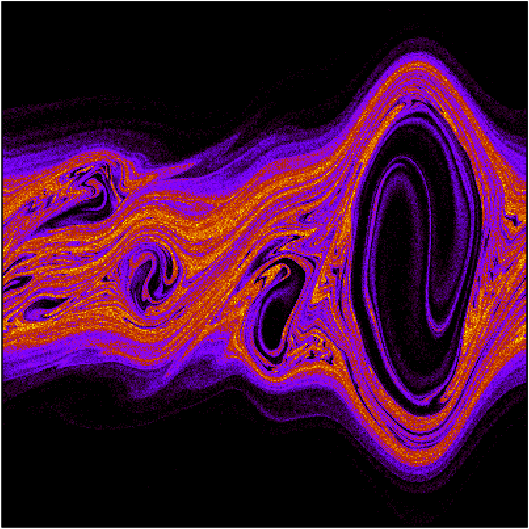} &
    \includegraphics[width = 0.10\textwidth]{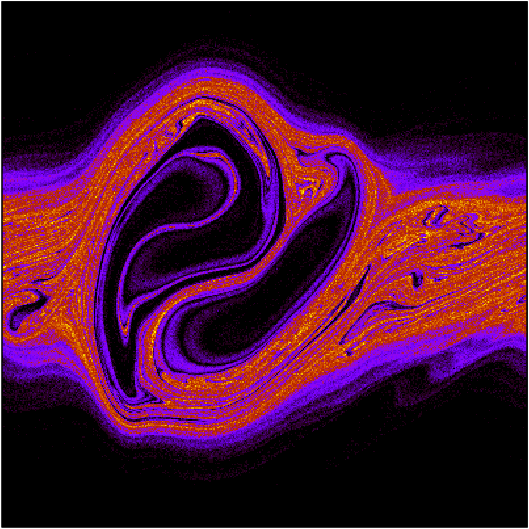} &
    \includegraphics[width = 0.10\textwidth]{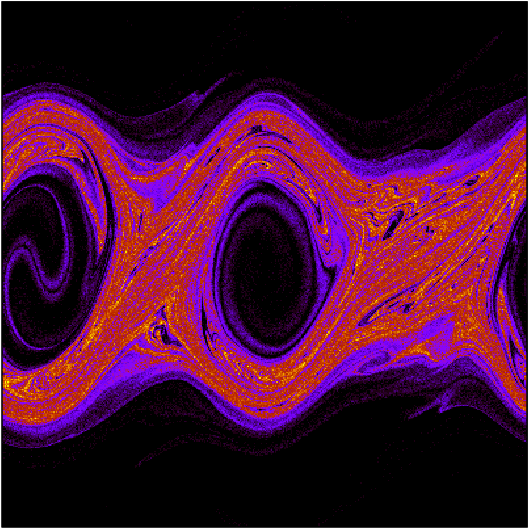} &
    \includegraphics[width = 0.10\textwidth]{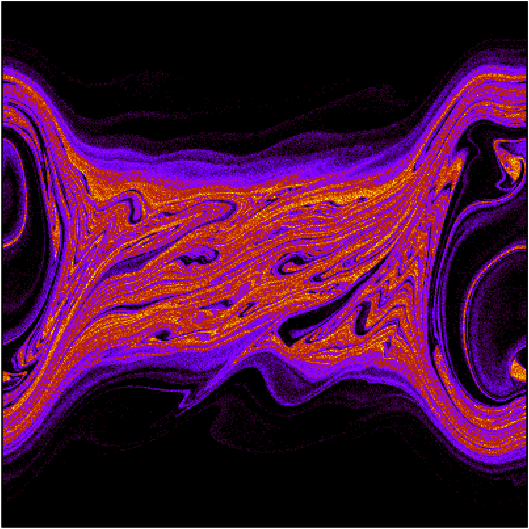} &
    \includegraphics[width = 0.10\textwidth]{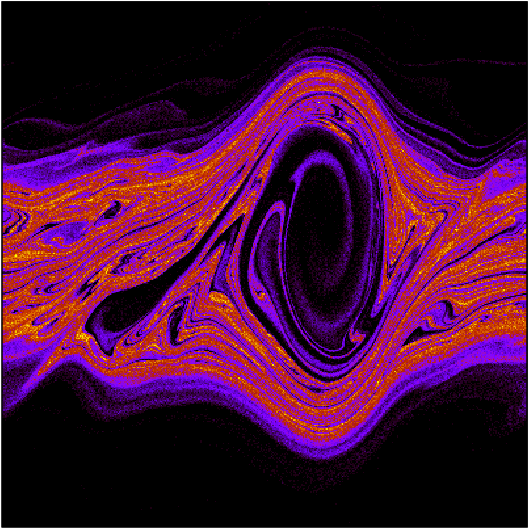} &
    \includegraphics[width = 0.10\textwidth]{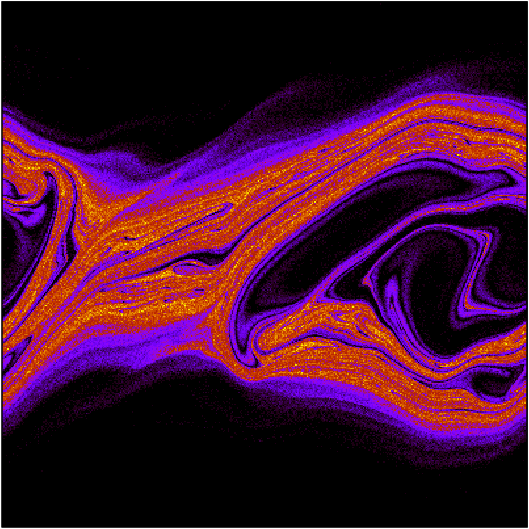} &
    \includegraphics[width = 0.10\textwidth]{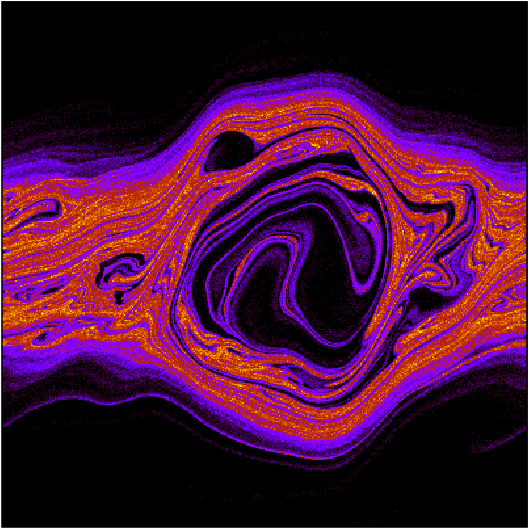} \\
    ($\varepsilon$) &
    \includegraphics[width = 0.10\textwidth]{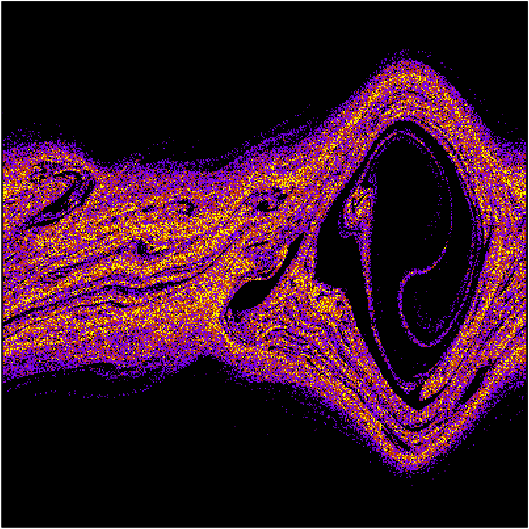} &
    \includegraphics[width = 0.10\textwidth]{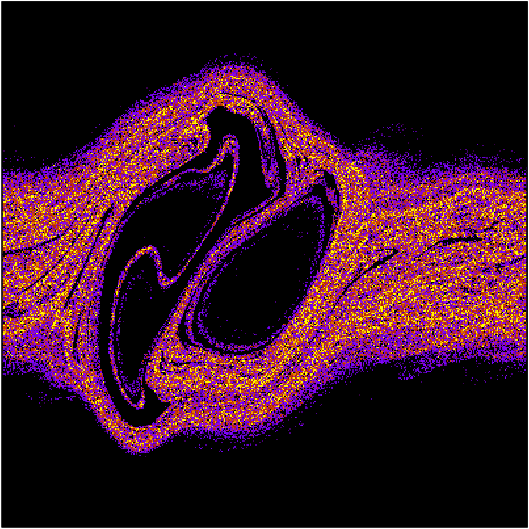} &
    \includegraphics[width = 0.10\textwidth]{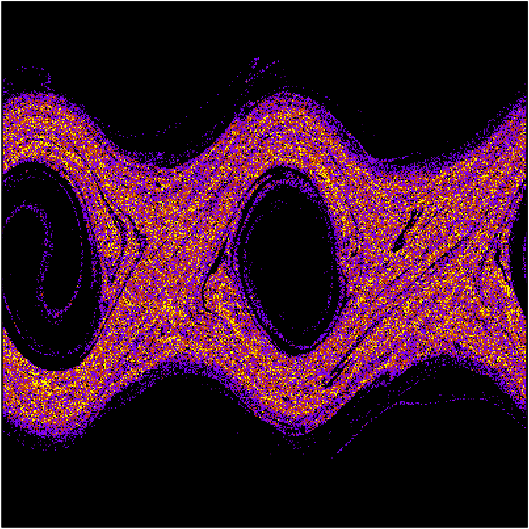} &
    \includegraphics[width = 0.10\textwidth]{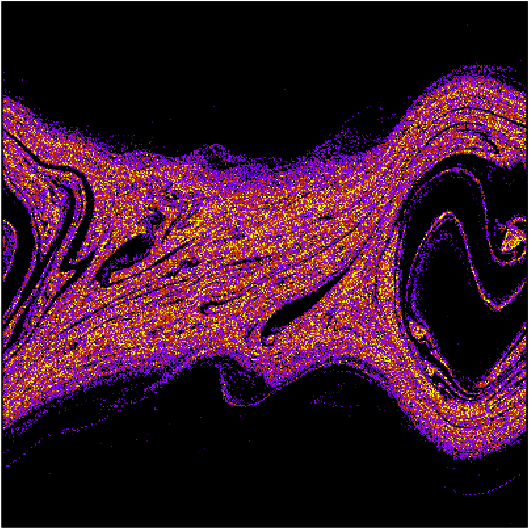} &
    \includegraphics[width = 0.10\textwidth]{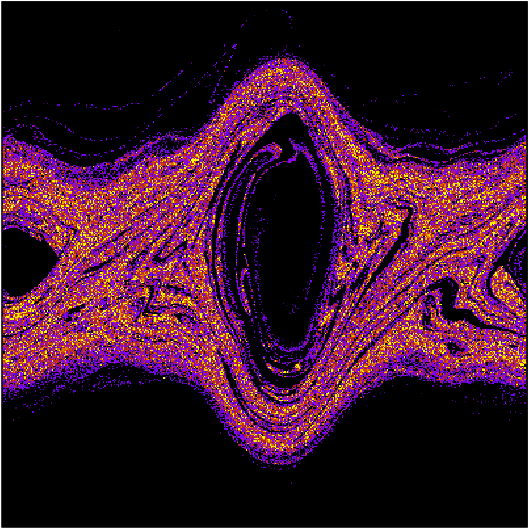} &
    \includegraphics[width = 0.10\textwidth]{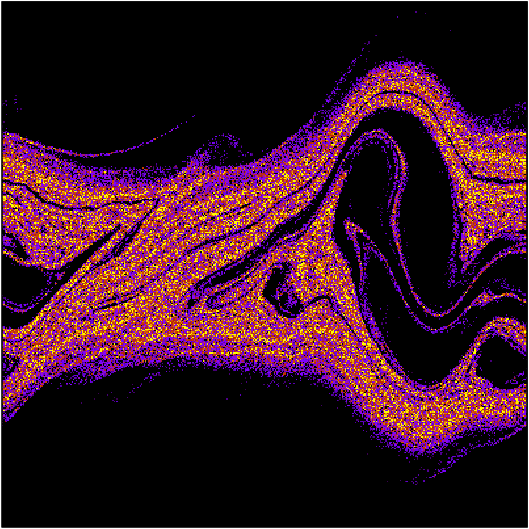} &
    \includegraphics[width = 0.10\textwidth]{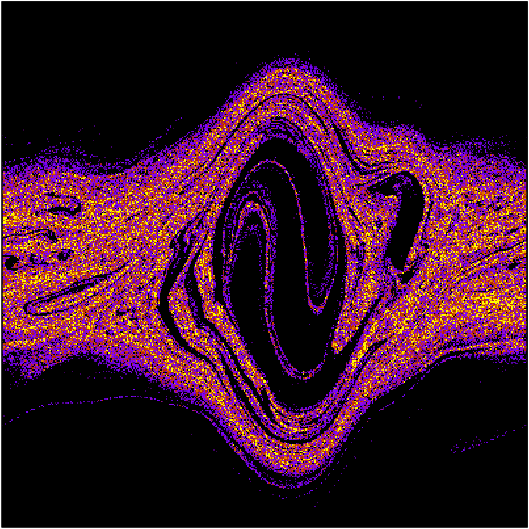} \\
    (p) &
    \includegraphics[width = 0.10\textwidth]{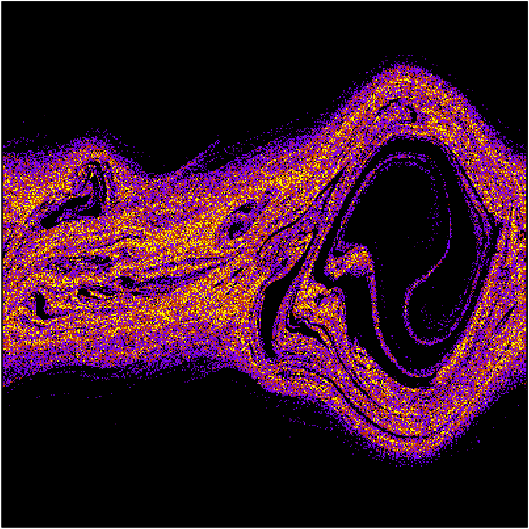} &
    \includegraphics[width = 0.10\textwidth]{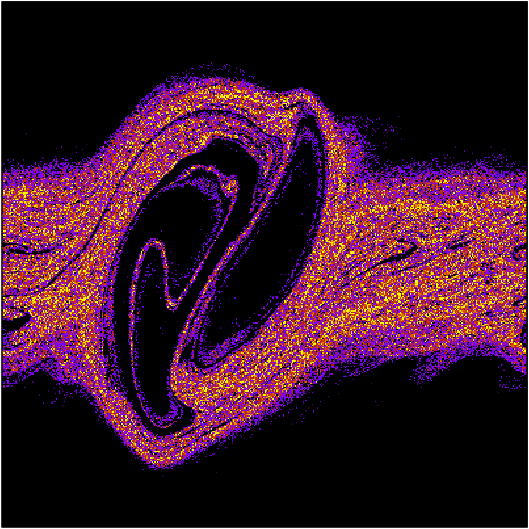} &
    \includegraphics[width = 0.10\textwidth]{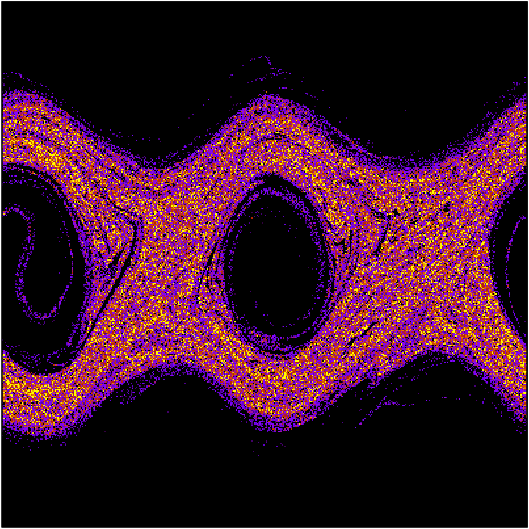} &
    \includegraphics[width = 0.10\textwidth]{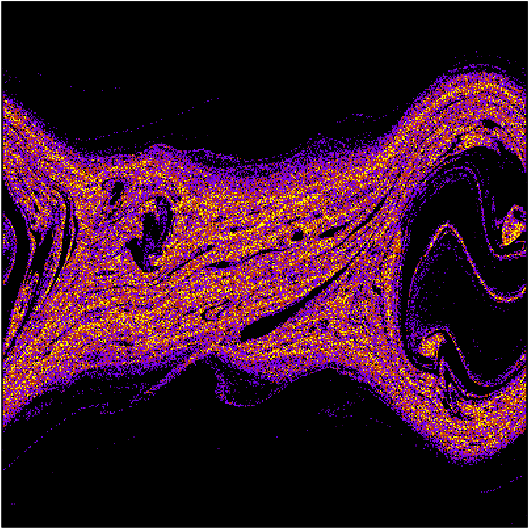} &
    \includegraphics[width = 0.10\textwidth]{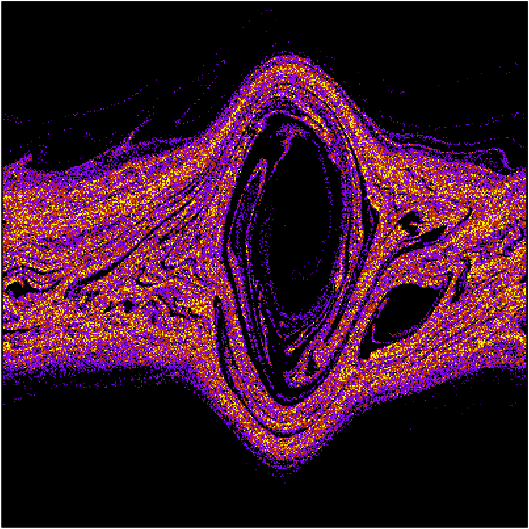} &
    \includegraphics[width = 0.10\textwidth]{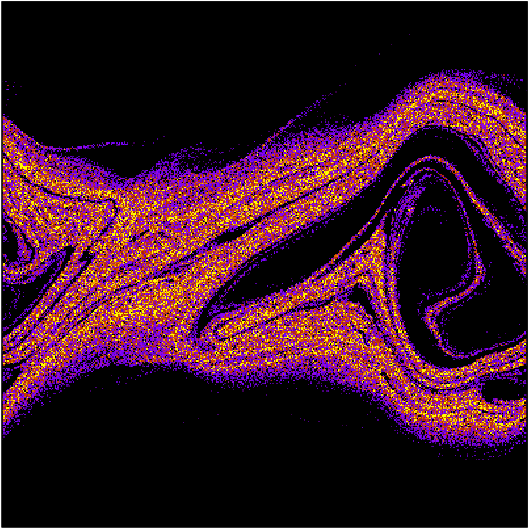} &
    \includegraphics[width = 0.10\textwidth]{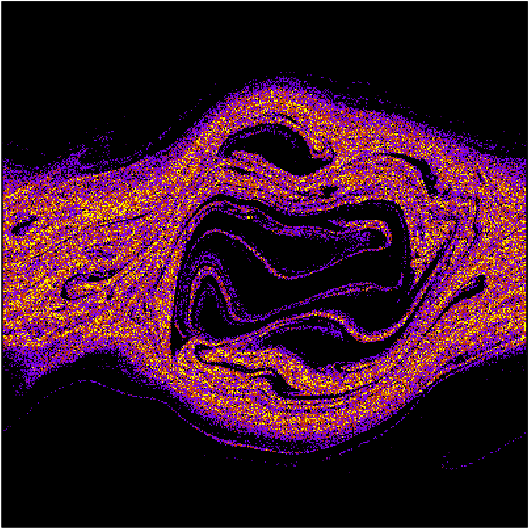} \\
    ($\vec{v}_r$) &
    \includegraphics[width = 0.10\textwidth]{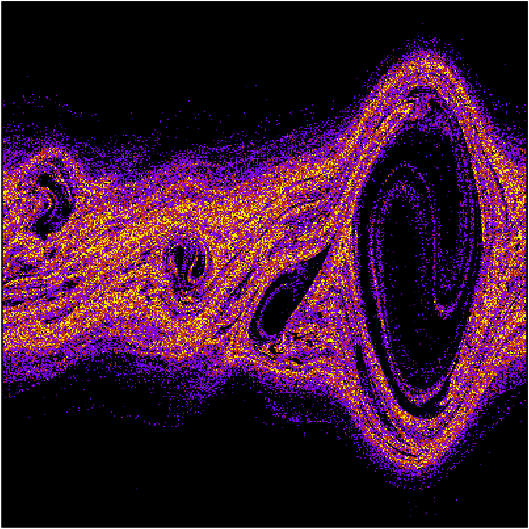} &
    \includegraphics[width = 0.10\textwidth]{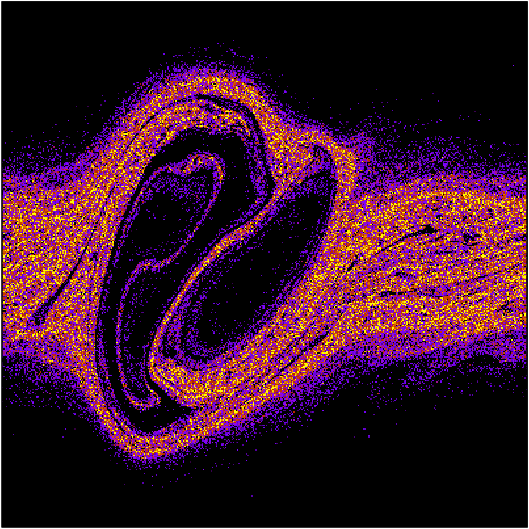} &
    \includegraphics[width = 0.10\textwidth]{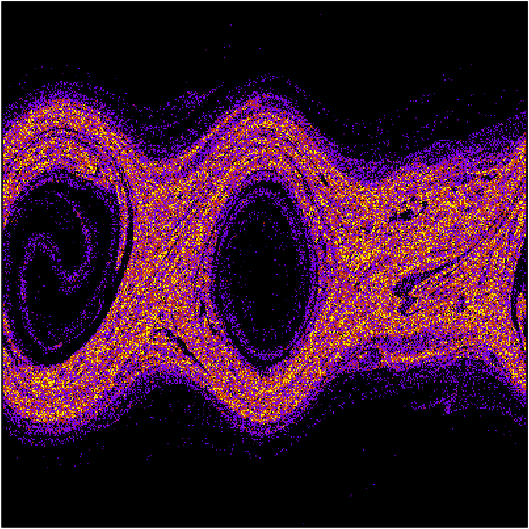} &
    \includegraphics[width = 0.10\textwidth]{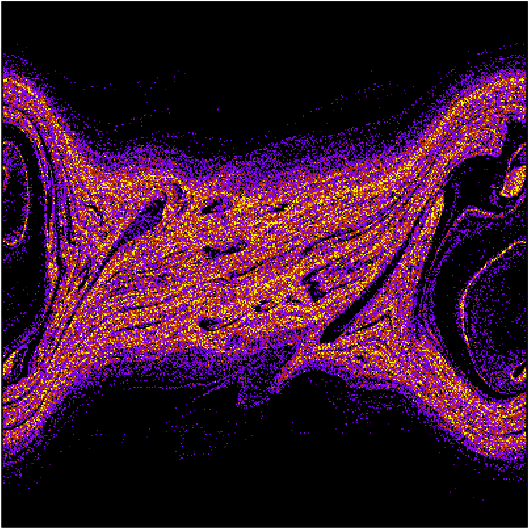} &
    \includegraphics[width = 0.10\textwidth]{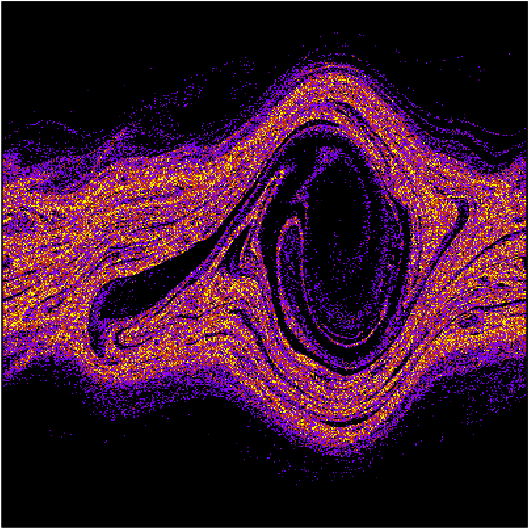} &
    \includegraphics[width = 0.10\textwidth]{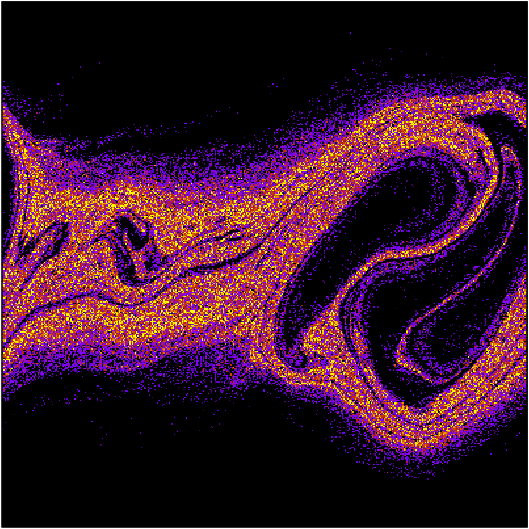} &
    \includegraphics[width = 0.10\textwidth]{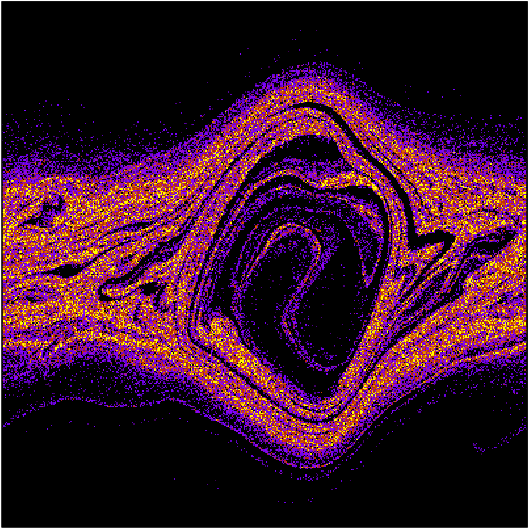} \\
    ($\vec{v}_r\varepsilon$) &
    \includegraphics[width = 0.10\textwidth]{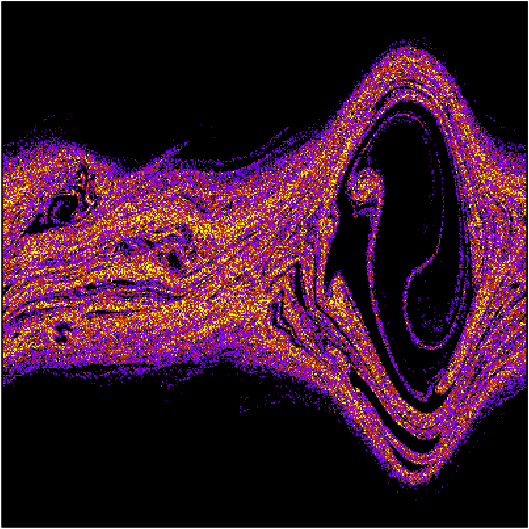} &
    \includegraphics[width = 0.10\textwidth]{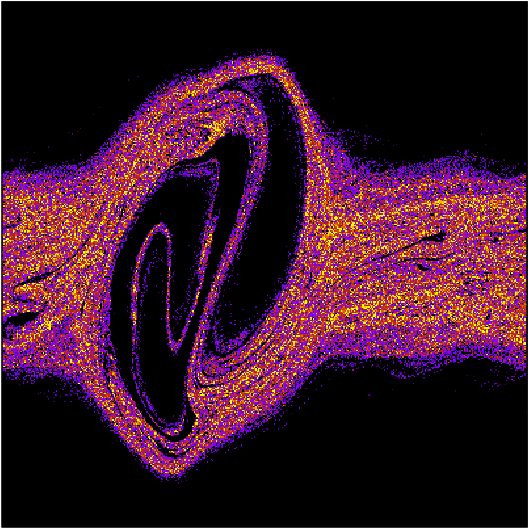} &
    \includegraphics[width = 0.10\textwidth]{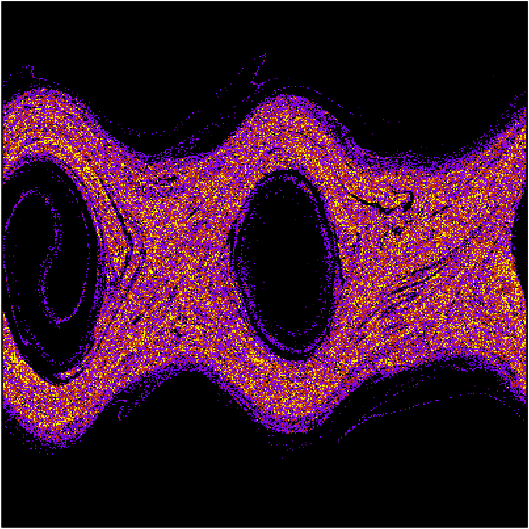} &
    \includegraphics[width = 0.10\textwidth]{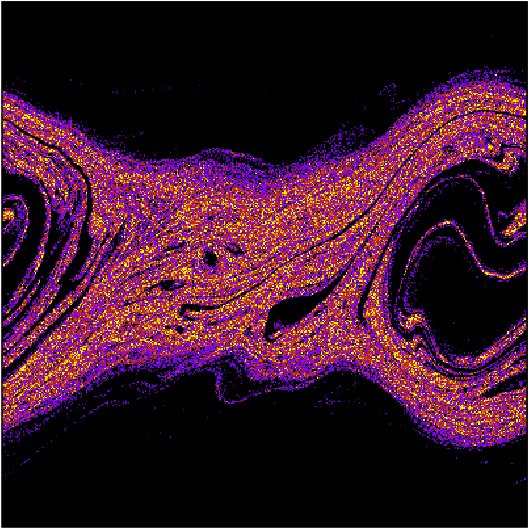} &
    \includegraphics[width = 0.10\textwidth]{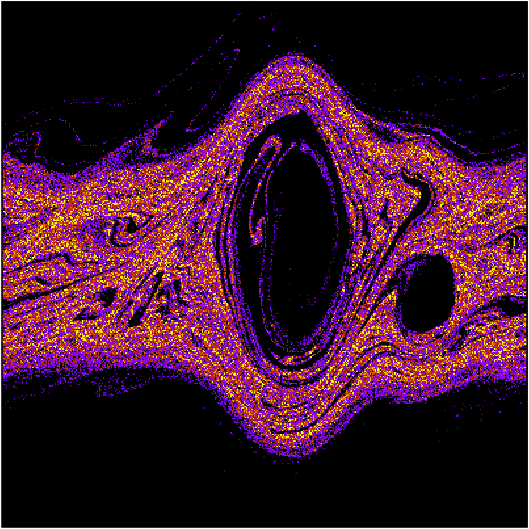} &
    \includegraphics[width = 0.10\textwidth]{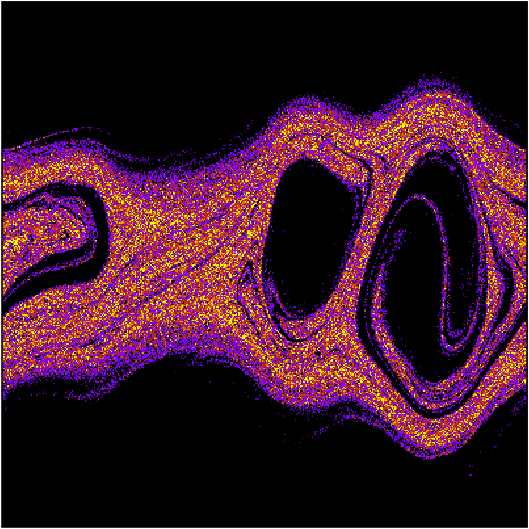} &
    \includegraphics[width = 0.10\textwidth]{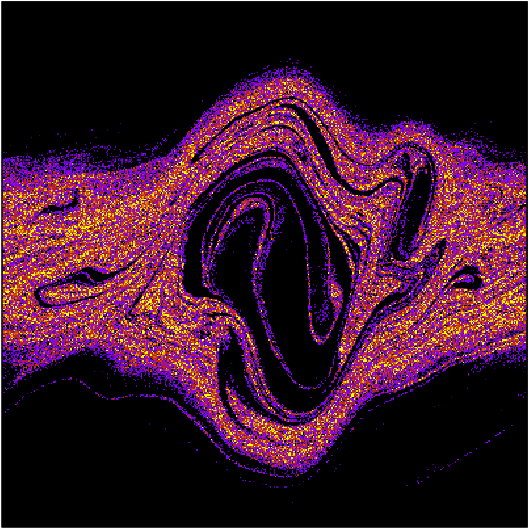} \\
    & run 1 & run 2 & run 3 & run 4 & run 5 & run 6 & run 7
  \end{tabular}
  \caption{Simulation results of the two-stream instability. This figure shows the same as figure~\ref{fig:two_str_merging_1}, but at $t = 90/\omega_p$ and
    with the merging done at $t = 20 / \omega_p$.
    Here, the $\vec{v}_r$ scheme seems to best preserve the physical evolution.
    \label{fig:two_str_merging_2}}
\end{figure}

How the physical evolution of the system is affected by the merging is shown in figures~\ref{fig:two_str_merging_1} and~\ref{fig:two_str_merging_2}.
Figure~\ref{fig:two_str_merging_1} shows simulation results at $t = 60 / \omega_p$, when merging was performed at $t = 5 / \omega_p$ and
figure~\ref{fig:two_str_merging_2} shows results at $t = 90/\omega_p$, for the case of merging at $t = 20 /\omega_p$.
In both figures, we show seven runs, that differ only in the initial state of the pseudorandom number generator.
After merging, there are about $8\cdot 10^4$ particles per beam, instead of the initial $10^6$.
The simulation results without any type of merging are also shown, for comparison.

\begin{figure}
  \centering
  \footnotesize
  \input{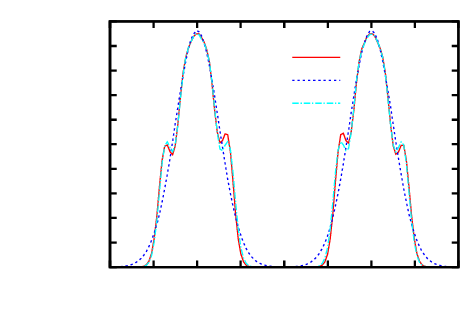}
  \caption{The velocity distribution function for various merge schemes,
    just after merging took place at $t = 5 /\omega_p$ in the two-stream simulation.
    The curves shown represent the average over 100 runs, to smooth out noise.
    In each run, the number of particles was reduced from $10^6$ to about $8\cdot 10^4$, by merging five times.
    The scheme $\vec{v}_r$ does on average not alter the velocity distribution, so its curve coincides with the curve before merging.
    The $\varepsilon$ and $\vec{v}_r\varepsilon$ scheme lead to almost indistinguishable velocity distributions after merging, and are therefore shown together.
    The p scheme gives similar results as these two:
    the tails of the distribution are moved towards the center, creating visible bumps in the velocity distribution function.
  }
  \label{fig:two_str_vdf_diff}
\end{figure}

\begin{figure}
  \centering
  \footnotesize
  \input{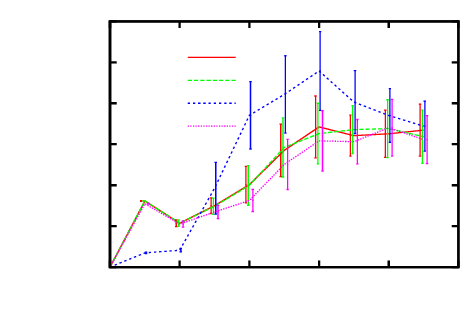}
  \input{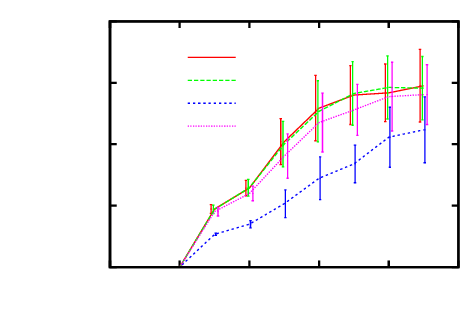}
  \caption{The difference in the velocity distribution function over time, caused by the various merge schemes in the two-stream simulation.
    The left figure shows results for merging at $t = 5/\omega_p$, while the right figure shows results for merging at $t = 20/\omega_p$.
    The quantity shown is the $L^2$ norm of the difference in the velocity distribution due to merging divided by the norm of the original velocity distribution,
    or $\vectornorm{\vec{f}_v(t) - \vec{f}_{v,0}(t)} / \vectornorm{\vec{f}_{v,0}(t)}$ as in equation~(\ref{eq:l2norm_vdf}).
    The error bars indicate the standard deviation in this quantity from run to run, computed from 100 different runs.
    The $\vec{v}_r$ scheme performs the worst for merging at $t = 5/\omega_p$, but when merging at $t = 20/\omega_p$ it performs the best.
  }
  \label{fig:two_str_vdf_diff_time}
\end{figure}

In figure~\ref{fig:two_str_vdf_diff}, we show the velocity distribution just after merging at $t = 5 /\omega_p$.
The $\vec{v}_r$ scheme does, on average, not alter the velocity distribution of the particles.
The other schemes all take averages in determining the properties of the merged particles, thereby removing the tails from the distribution.
However, the effect a merge scheme has \emph{on average} says little about the fluctuations it introduces.
Therefore we also include figure~\ref{fig:two_str_vdf_diff_time},
in which the difference in the velocity distribution as compared to the original simulation is shown over time.
More precisely, we show the quantity
\begin{equation}
  \vectornorm{\vec{f}_v(t) - \vec{f}_{v,0}(t)} / \vectornorm{\vec{f}_{v,0}(t)},
  \label{eq:l2norm_vdf}
\end{equation}
where $\vec{f}_v(t)$ and $\vec{f}_{v,0}(t)$ denote the `merged' and the original velocity distribution function, respectively, that were constructed using 200 bins.
(As before, we use $\vectornorm{}$ to indicate the $L^2$ norm.)

From figures~\ref{fig:two_str_merging_1} and~\ref{fig:two_str_vdf_diff_time}, it can be seen that when merging happens at $t = 5 / \omega_p$, the momentum conserving scheme (p) seems to perform best.
The schemes that conserve energy ($\varepsilon$ and $\vec{v}_r\varepsilon$) perform almost as good, but the scheme that picks velocities at random ($\vec{v}_r$) does considerably worse.
The reason for this is probably that the $\vec{v}_r$ schemes induce greater fluctuations in the momentum distribution, see table~\ref{tab:schemetable}.
When the two-stream instability has a small magnitude, these induced fluctuations perturb the further evolution of the system.

For the case of merging at $t = 20 / \omega_p$ the results look quite different, see figures~\ref{fig:two_str_merging_1} and~\ref{fig:two_str_vdf_diff_time}.
Now, the $\vec{v}_r$ scheme has the smallest effect on the evolution of the simulation, with the other three schemes performing worse.
The larger induced fluctuations of the $\vec{v}_r$ scheme are probably not as important now,
because the two-stream instability has already grown to a larger size before merging.

In the previous sections, we have also presented results for $k$-d trees that used the norm of the velocity vector.
In one dimension, this corresponds to taking the absolute value of the velocity.
But because in the two-stream simulation particles flow in two opposite directions, this leads to poor results.
Particles from the two beams are randomly mixed, significantly altering the velocity distribution.
In general, using the norm of the velocity vector should probably only be used when the flow of particles is in a single direction.

When we perform the merging cell-by-cell, the results show no clear differences from those presented in figures~\ref{fig:two_str_merging_1} and~\ref{fig:two_str_merging_2}.

\subsection{Computational costs of $k$-d trees}
\begin{figure}
  \centering
  \begin{minipage}{0.49\textwidth}
    \centering
    \footnotesize
    \input{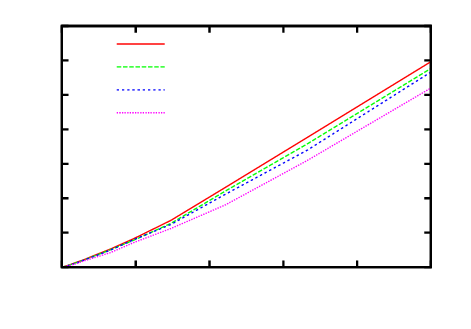}
  \end{minipage}
  \begin{minipage}{0.49\textwidth}
    \centering
    \footnotesize
    \input{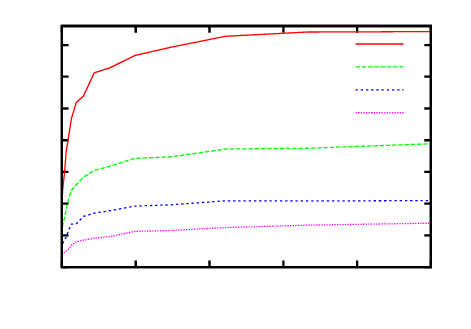}
  \end{minipage}
  \caption{Performance figures for $k$-d trees in 2D-5D with $N_\mathrm{p}$ points, using the \texttt{KDTREE2}~\cite{Kennel8067K} library.
    Left: the time it takes to create the $k$-d tree.
    Right: the time it takes to find a nearest neighbor (for uncorrelated searches).
    The calculations were performed on an Intel i7-2600 CPU.
  }
  \label{fig:kdtreeperf}
\end{figure}

The goal of an APM algorithm is to speed up a simulation, so the algorithm itself should not take too much time.
Theoretically, the computational complexity of creating a $k$-d tree is $O(N_\mathrm{p} \log N_\mathrm{p})$, with $N_\mathrm{p}$ the number of points in the tree.
The average cost of a random search in the tree is $O(\log N_\mathrm{p})$.
We have tested the practical performance of the \texttt{KDTREE2} library on an Intel i7-2600 CPU.
In figure~\ref{fig:kdtreeperf} the creation time and the average search time are shown for $k$-d trees of various sizes.
Neighbors can be found faster if the $k$-d tree is constructed in fewer dimensions.
Note that the average search time is given for uncorrelated searches, that are done at random locations.
This is the worst-case scenario, as the CPU cannot do efficient data caching.
If the next search location is picked close to the previous search location, search times in 5D decrease by more than $80\%$.

The time scales for constructing and searching a $k$-d tree can be compared,
for example, with the cost of updating a particle in an electrostatic plasma simulation with collisions.
On the same machine, about 0.1--1 $\mu$s is spend per particle on interpolating forces from the grid,
updating the particle position and velocity, determining whether a collision should occur,
mapping the particles to densities again and computing the electric field.
In such simulations merging would typically not occur at every timestep, and only for a fraction of the particles.
Therefore, the computational cost of setting up $k$-d trees and searching for neighbors would not contribute much to the simulation time.

If in a simulation the cost of advancing particles is very small, but their weights have to adjusted very often, then the use of $k$-d trees might slow the simulation down.
In such cases, it might be better to divide the particles over the grid cells, which can be done much faster than setting up a $k$-d tree, and then use a fast algorithm that operates on a cell-by-cell basis to adjust the weights.

\section{Conclusion}
Adaptively adjusting the weights of simulated particles can greatly improve the efficiency of simulations.
We follow Welch et al.~\cite{Welch2007143} and call algorithms that do this `adaptive particle management' (APM) algorithms.
In this work, we have focused on the pairwise merging of particles.
We found that the use of a $k$-d tree offers several important advantages over present methods.
First, only particles that are `close together' are merged.
`Close together' can be defined as desired (for example close in position and velocity).
This ensures that the distribution of particles is not significantly altered.
Second, the merging can be performed completely independent of the numerical mesh used in the simulation.
The algorithm works in the same way, whether the simulation is in 1D or in any higher dimension.
Third, with a $k$-d tree, the closest neighbors can be located efficiently.
Therefore, the method can be used for simulations with millions of particles.
Fourth, from a practical point of view, the use of a $k$-d tree library greatly simplifies the implementation of pairwise merging.

Two particles can be merged in different ways, and we have compared various merge schemes.
An interesting option is to select properties for the merged particle at random from the original particles.
With these stochastic schemes fluctuations increase, but on average both momentum and energy can be conserved.
In the simulation of the two-stream instability we saw that when system is sensitive to small fluctuations, a scheme that conserves energy or momentum is preferred.
But when the physical evolution does not depend strongly on small fluctuations, a stochastic scheme can perform better.

In general, it is more important to preserve the essential characteristics of the particle distribution function than to exactly conserve grid moments.
A scheme that conserves energy or momentum should typically be used with a full coordinate $k$-d tree (containing $\vec{x},\vec{v}$).
A velocity norm $k$-d tree (containing $\vec{x},\vectornorm{\vec{v}}$) can be used with a stochastic scheme.
The advantage of a velocity norm $k$-d tree is that is can be constructed and searched faster than one with the full coordinates.
The combination of a stochastic scheme with a full coordinate $k$-d tree seems a good choice:
on average, the shape of the energy and momentum distribution functions is conserved, while the induced fluctuations in the grid moments are relatively small.

\section*{Acknowledgement}
We would like to thank both referees for their comments, that significantly improved this article. J. Teunissen was supported by STW-project 10755.

\bibliographystyle{elsarticle-num.bst}
\bibliography{particle_control.bib}

\end{document}